\documentclass[preprint,3p,12pt,authoryear]{elsarticle}
\usepackage{babel}
\usepackage{graphicx, color}
\usepackage{bbm}
\usepackage{setspace}
\usepackage{amsmath,amssymb}
\usepackage{amsfonts}
\usepackage[export]{adjustbox}
\usepackage{float}
\usepackage[hyphens]{url}
\PassOptionsToPackage{hyphens}{url}\usepackage{hyperref}
\usepackage{subcaption}
\usepackage{mwe}

\DeclareMathOperator*{\argmax}{argmax}
\newcommand{\argmin}{\operatornamewithlimits{argmin}}
\renewcommand{\vec}[1]{\mathbf{#1}}
\newcommand{\bphi}{\mbox{\boldmath $\phi$}}
\newcommand{\bxi}{\mbox{\boldmath $\xi$}}

\newcommand{\bmu}{\mbox{\boldmath $\mu$}}

\usepackage{amsthm}
\usepackage{xcolor}

\newtheorem{theorem}{Theorem}[section]
\newtheorem{definition}[theorem]{Definition}

\graphicspath{ {./figures/} }

\begin{document}

\newcommand{\yuri}[1]{{\color{orange} #1}} 
\newcommand{\brenda}[1]{{\color{violet} #1}} 
\newcommand{\com}[1]{{\color{red} #1}} 
\newcommand{\final}[1]{{\color{black} #1}}

\begin{frontmatter}



\title{Functional Classification of Bitcoin Addresses}


\author[1]{Manuel Febrero-Bande}
\ead{manuel.febrero@usc.es}

\author[1]{Wenceslao Gonz\'alez-Manteiga}
\ead{wenceslao.gonzalez@usc.es}

\author[2]{Brenda Prallon}
\ead{bq45@cornell.edu}

\author[3]{Yuri F. Saporito}
\ead{yuri.saporito@fgv.br}

\address[1]{Departamento de Estad\'istica, An\'alisis Matem\'atico y Optimizaci\'on, Universidade de Santiago de Compostela, Spain}
\address[2]{Department of Economics, Cornell University, United States}
\address[3]{Escola de Matemática Aplicada, Fundação Getulio Vargas, Brazil}

\begin{abstract}
This paper proposes a classification model for predicting the main activity of bitcoin addresses based on their balances. Since the balances are functions of time, we apply methods from functional data analysis; more specifically, the features of the proposed classification model are the functional principal components of the data. Classifying bitcoin addresses is a relevant problem for two main reasons: to understand the composition of the bitcoin market, and to identify \final{addresses} used for illicit activities. Although other bitcoin classifiers have been proposed, they focus primarily on network analysis rather than curve behavior. Our approach, on the other hand, does not require any network information for prediction. Furthermore, functional features have the advantage of being straightforward to build, unlike expert-built features. Results show improvement when combining functional features with scalar features, and similar accuracy for the models using those features separately, which points to the functional model being a good alternative when domain-specific knowledge is not available.

\end{abstract}




\end{frontmatter}


\section{Introduction}
\label{sec:introduction}

This work tackles the question of building a model to classify the main activity of bitcoin addresses. Bitcoin was the first decentralized cryptocurrency to be created, and it is also the most popular. Anonymity is a central characteristic of the bitcoin protocol, and one of the reasons for its success. This implies, nevertheless, that the market composition of this cryptocurrency is not obvious: the purposes for which bitcoins are spent are obscure. It is known, however, that illicit services account form a relevant portion of that market\footnote{46\% of bitcoin transactions, as estimated by Foley et al. (2019).} - which follows from anonymity certainly being an attractive for law-breakers\footnote{The Silk Road darknet market, closed by the FBI in 2013 and used mainly for commercializing illegal drugs, moved approximately fifteen million dollars annually in transactions (Christin, 2013).}. Thus, identifying the main activity of a bitcoin address is a relevant issue, since it both aids law enforcement and sheds some light on the bitcoin market organization. 

\subsection{\final{Literature Review on Bitcoin Addresses' Classification }}

Regarding the classification of bitcoin data, most of the work focuses on identifying illicit activities, and uses some type of aggregation of \final{addresses}, linking them to one common entity. Furthermore, the literature tends to utilize features from network relations, instead of the functional behavior of \final{transactions}. \cite{meiklejohn2013fistful} propose a method to cluster bitcoin addresses to user-level; that is, multiple addresses are assigned to each user (also called entity). They use a small number of transactions labeled through their own empirical interactions with various entities, and then identify major institutions and their interactions. 

\cite{foley2019sex} study the bitcoin market of illicit activities. They find that illegal users of bitcoin tend to transact more, in transactions involving fewer addresses, and they tend to hold smaller amounts of the cryptocurrency. To make this analysis, they first cluster addresses to user-level using the approach from \cite{meiklejohn2013fistful}. Then, two models are developed: the first one used information on the addresses network - that is, which addresses communicated with each other - to cluster legal/illegal activities. The second method is detection-controlled estimation. The covariates used account for transaction frequency, USD value, and \final{addresses'} lifespan, but there are also other variables with information of the network, such as how many users are involved in the transactions, and information of external shocks. 

Graph neighborhood features were proposed by \cite{jourdan2018characterizing} to classify users with five different labels, taken from \texttt{WalletExplorer}\footnote{\url{walletexplorer.com}}: exchange, gambling, mining, service and darknet. They build a total of 315 features, but obtain great accuracy results using only 15. They show that graph features improve the accuracy significantly over features considering only the entities addresses. Logistic regression and gradient boosting were employed, and results show that the darknet is perfectly classified, and other classes, apart from mining, also score well. \cite{hu2019characterizing} also propose graph features to identify money laundering activities occurring across the bitcoin network. Their binary model classify transactions, with high accuracy.

\subsection{\final{Literature Review on FDA}}

There are many specific methods for classifying functional data. In this article, the focus is on adapting classifiers from the multivariate framework.
This is achieved by projecting the curves of \final{credits and debits} onto a functional basis, and then using the coefficients as functional features.

The proposed basis is the eigenbasis of the Karhunen-Loève expansion, the functional principal components (FPCs) basis. Besides being widely explored in the literature, this basis maximizes the variance of the observed curves, analogously to classical multivariate principal components; this means that, typically, few components are necessary to explain most of the differences of the functions;  \cite{hall2001functional} argue that the FPCs capture the greatest part of the curves, in a $L^2$ sense, since they are maximizing variance. They apply this transformation to radar signals curves, and estimate the coefficients density to compute the posterior probabilities in order to classify the signals in eight different groups. Their model performs significantly better than the multivariate counterpart. 

In \cite{muller2005generalized} it is shown that, when using the principal components basis to represent the functions and truncating the expansion, the functional logistic model becomes the equivalent of a multiple logistic regression with FPCs. \cite{escabias2004principal} compare two different approaches for a functional logistic model. The first one consists of smoothing the curves on an arbitrary basis, and then performing regular PCA on the matrix defined by the multiplication of the coefficients matrix by the inner product matrix. The second one consists on estimating the FPCs by the method described in Section 8.4.2 of \cite{RamsaySilverman2005book}. The resulting principal components are inputs for the multiple logistic regression. In \cite{escabias2005modeling}, the first method is applied with B-splines to predict risk of drought across different areas in Canada.  

A functional logistic model is also used to classify gene expression data in \cite{leng2006classification}. On the same subject, \cite{song2008optimal} develop a model to classify gene expression that consists of applying multivariate classification algorithms to FPCs estimated from the data. They first smooth the data with a pre-defined basis and then compute the principal components. The pre-defined basis is chosen as a cubic B-spline basis, since it is flexible and computationally efficient.

Following the same idea, \cite{li2013hyperspectral} attempt to classify hyperspectral images by treating their pixels as curves instead of high-dimensional discrete vectors. As in \cite{song2008optimal}, the curves were first smoothed using cubic B-splines, with the difference that, instead of choosing the number of basis, they use a smoothing parameter for penalizing the second derivative; functional principal components were computed in the same fashion, and the number of FPCs was chosen by five-fold cross-validation. We use the same procedures in our work. The method was compared to other methods, including Support-Vector Machine (SVM) applied to regular principal components, on three popular hyperspectral data sets. Results show that their approach performs consistently better. Functional principal components as features for SVMs were also employed in \cite{lee2004functional}.

Furthering the topic of FPCs, \cite{hall2006properties} derive some asymptotic bounds for the truncated estimators of eigenfunctions and eigenvalues. \cite{yao2005functional} propose a nonparametric method to perform functional principal components analysis (FPCA) with sparse logitudinal data. Finally, since FPCA is very sensitive to outliers, some tools to identify and remove those based on notions of functional data depth have been discussed in \cite{lopez2007depth} and \cite{cuevas2007robust}.

\final{Additionally to the classical reference \cite{RamsaySilverman2005book}, we forward the reader to some recent reviews and new methods on functional classification methods using FDA: \cite{febrero2013generalized}, \cite{jiang2020filtering}, \cite{aneiros2021variable}, \cite{piotr2017special}, \cite{ling2021semiparametric} and \cite{li2022multivariate}, \cite{wang2016functional}}.

\subsection{\final{Contribution}}

We propose a solution to \final{the classification of bitcoin addresses} considering information of addresses' balances. More precisely, we use the \final{addresses' transactions}, which are the credits and debits made throughout a period of time. This is essentially a task of classification by analyzing the behavior of curves. Although, \final{as we have aforementioned,} there have been a few works with the same ultimate goal - to classify bitcoin addresses by activity -, the classifiers in the literature focus on arbitrary features, that come from field specific knowledge - also known as the process of `feature engineering'- or network analysis. Our approach differs from these because we use the fact that the account movements are a function of time, or, more specifically, of the address' life span \footnote{Here understood as the time elapsed between the first credit, at $t = 0$, up to a pre-defined limit, $t = T$.}. That is, we aim to find patterns in the shapes of the \final{credits and debits} curves, and to classify the \final{addresses} based on them. This is done by employing Functional Data Analysis (FDA) tools. The advantage of this strategy is, beyond its straightforwardness, that we do not need to have any information about which addresses are communicating with each other on our training set in order to make predictions, which is very useful for \final{classifying} addresses mainly connecting to unlabeled vertices of the network; \final{this is our main contribution}.

% \com{add A comment about the difficulty of the task given that the balances do not characterize the type of activity or even the activity could be of mixed nature. Another difficulty in the imbalances of the problem}

The arbitrary features normally contain implicit information of some aspects of curve behavior; for example, the number of credits is a measure for frequency, and the total amount of credit is a measure of level. FDA, however, accomplishes summarizing this information in a more direct way, meaning that the methods do not require as much domain-specific knowledge. This is particularly useful when the process of feature engineering is complicated.

This paper is organized as follows: Section \ref{sec:empirical_strategy} discusses data acquisition and treatment, including details on how to represent the data as functional. Section \ref{sec:classification_models} discusses the different types of features considered, both scalar and functional. Section \ref{sec:results} presents the results, and Section \ref{sec:conclusion} concludes the paper.

\section{Empirical Strategy}
\label{sec:empirical_strategy}

\subsection{Data Acquisition}

\final{A bitcoin address is a unique string of number and letters stored in the bitcoin's blockchain that can be the recipient or sender of bitcoins. A bitcoin transaction can have multiple addresses as inputs and outputs. One could think the address as the number of a bank account, but with some very distinct properties as public visibility of balance and all transactions, anonymity, cannot hold negative quantities of bitcoins (debt), and each user in this market usually has a large number of addresses. The user here, though, is not an actual individual, but a company/website/organization that typically holds multiple addresses; we refer to it as ``entity".
The balance of an address is the amount (possibly zero) of bitcoins in that address at a given time. The entity that a given address belongs to is usually unknown but some companies make some of its addresses public for various reasons. Based on this public information, one can identify more addresses as belonging to this same entity if they were inputs to the same transaction with one or more inputs from this entity.}

The data used in this project includes both information on the address's class and their balances over time. This allows for the development of a supervised classification model in which the balance values are used to predict the class of a certain address, that is, a model which predicts the main activity of a given address based on its transactions.
Thus, the data was collected in essentially two stages: classifying the addresses by their major purpose and fetching their balances over time.

Through the scraping of \texttt{WalletExplorer} we obtained information linking addresses with entities (for example, address \texttt{1PkJRQaKStcmCeh} \texttt{CJUjxYiQejFDm3w4yV} belongs to \url{999Dice.com}). We then used the \final{classes} of \texttt{WalletExplorer}: exchange, gambling, pools, services/others, and old/historic; however, we only kept the addresses of the old/historic category related to illegal activities, and renamed the \final{class} ``darknet". This work focuses on linking addresses to these \final{classes} rather than entities. The addresses gathered were active at some time between April 2011 and April 2017. Furthermore, we used Google's BigQuery public dataset on bitcoins \footnote{\url{https://cloud.google.com/blog/products/data-analytics/introducing-six-new-cryptocurrencies-in-bigquery-public-datasets-and-how-to-analyze-them}} to obtain the \final{addresses' balances}.

The initial data set had the descriptions shown in Table \ref{n_obs_cat_orig}.

% latex table generated in R 3.6.3 by xtable 1.8-4 package
% Sat Apr 17 19:30:05 2021
\begin{table}[ht]
\centering
\begin{tabular}{lrr}
  \hline
Category & Addresses & Entities \\ 
  \hline
Darknet Marketplace & 106,274 &   8 \\ 
  Exchanges & 259,102 &  50 \\ 
  Gambling & 45,986 &  17 \\ 
  Pools & 56,811 &   1 \\ 
  Services/others & 287,160 &  22 \\ 
  Total obs. & 755,333 &  98 \\ 
   \hline
\end{tabular}
\caption{Original Number of Addresses and Entities, by Category} 
\label{n_obs_cat_orig}
\end{table}

\subsection{Data Treatment}

In this section, we describe all the treatment necessary to pose a well-defined classification problem and build the functional features. We choose a time frame to observe the addresses, creating a clear rule of how much time is necessary to wait after the first observed transaction before classifying them. We also limited the sample to a minimum of observations, in order to properly capture curve behavior. Measures of level \final{(shape of credits and debits)}, variation \final{(the derivative of credits and debits)} and frequency \final{(Poisson rate in which transactions occur)} are created using FDA methods, and their FPCs are computed. We used the R software, with packages \cite{fdauscpackage} and \cite{ramsay2020package}.

\subsubsection{A Sampling Issue}
\label{sec:sampling_issue}
The classification of bitcoin addresses by their balances over time depends on the underlying hypothesis that an address' behavior is invariant over \final{calendar} time. That is, for addresses of the same \final{class}, if one is observed in a year and another on the next, they should still present the same characteristics of movements and amounts. We argue that this amount should be measured in dollars. Since the bitcoin price is very volatile, if the balances were measured in bitcoins, they might impose a change in level through time. Again, for two addresses observed through different time frames, if the bitcoin price is higher in one period, they will probably be receiving smaller amounts of bitcoins for providing the same service. 

This difference in level of bitcoin amounts could possibly add a timestamp bias: the model may be classifying better the category with addresses observed in similar dates. If the samples of a certain category are concentrated around a period of time, say, beginning in March 2015, the bitcoin price volatility would affect them equally. If instead the samples of another category are addresses whose start is somewhere in the range from 2011 to 2017, the amount of bitcoins will most likely vary greatly between them, even if they present the same behavior. Greater variability due to differences in the start date of the samples of a category could make defining a pattern harder, thus harming the classification of that category.

\begin{figure}[H]
    \centering
     \includegraphics[scale = 0.6, center]{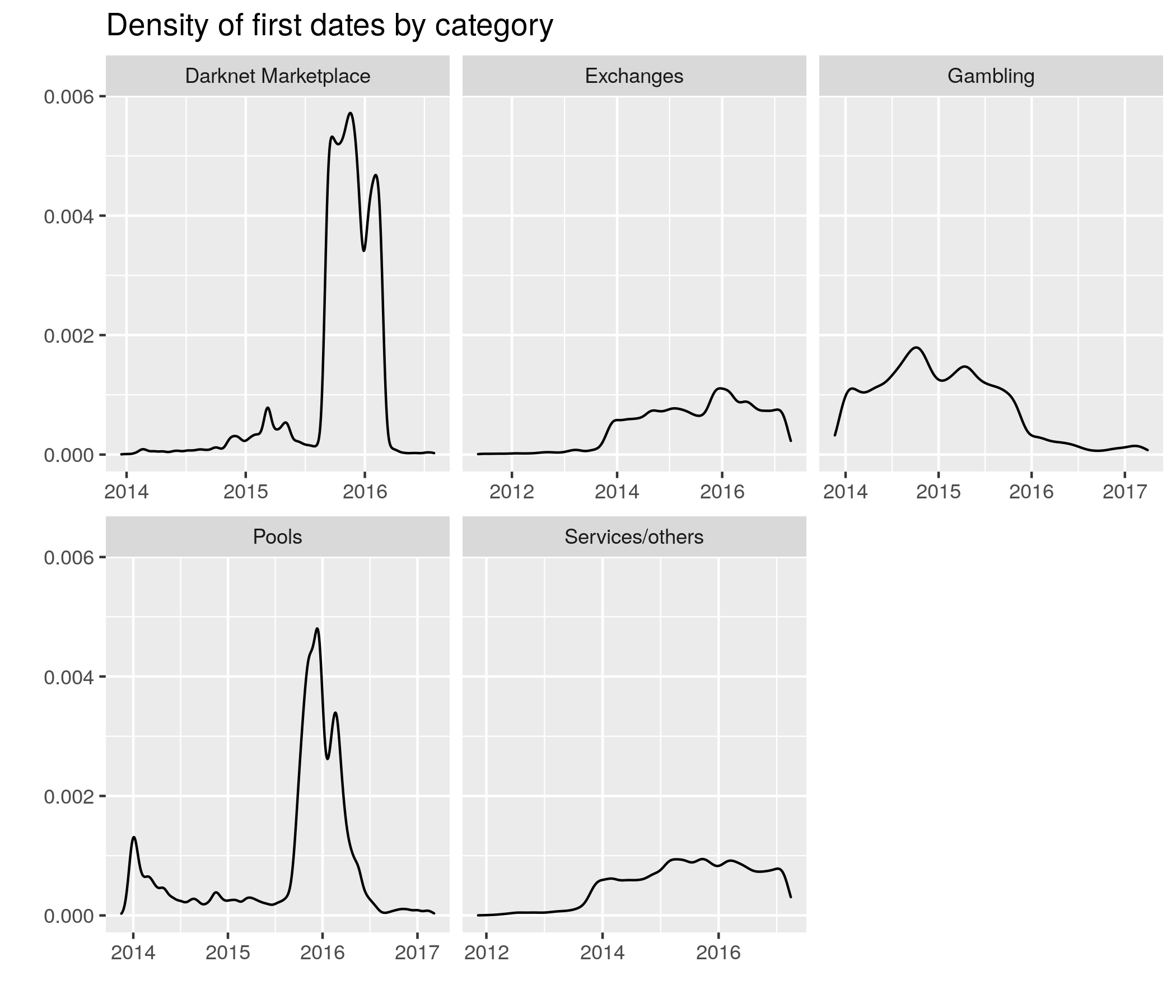}
     \caption{Distribution of starting dates by category.}
     \label{fig:cat_dates}
\end{figure}

Indeed, the addresses are not observed uniformly over time: \final{in Figure \ref{fig:cat_dates}, it is evident} that the darknet and pool samples are much more concentrated than their counterparts. This is due to the sampling nature: for darknet addresses to be identified and labeled, a police operation was most likely necessary - as was the case with the Silk Road website, shut down by the FBI in 2013. This poses another matter, that was not addressed in this work: services that put a lot of effort into anonymity, such as the darknet, will most likely have few entities discovered. This is in fact the case and Table \ref{n_obs_cat_orig} shows the imbalance \final{among the categories for the classification task.}

However, even with the values in dollars, there could still be an implicit timestamp in volatility: addresses in a time of high (or low) volatility could be grouped together. Furthermore, volatility introduces noise between one \final{transaction} and another. To avoid these phenomena, instead of transforming each amount by the bitcoin price at end of the day, the mean of the bitcoin price is taken over the period that the addresses are observed and then used to level the units. Therefore, the addresses balances are represented in dollars, instead of bitcoins. More specifically, the transformed balances are the balances in bitcoins multiplied by a fixed amount, which is the mean of the bitcoin value in dollars over the address lifespan.

\subsubsection{Functional Data: Additional Treatment}

It was required to limit the sample for addresses with a minimum number of observations - otherwise there is not enough points to treat the data as functional. This was tested using thresholds of 10 and 20 \final{transactions}. No higher threshold was tested because, for a minimum of 30 \final{transactions}, we only have fewer than 500 addresses classified as darknet, and we would have to keep a very small sample.  

\begin{figure}[H]
    \centering
    \includegraphics[scale = 0.6, center]{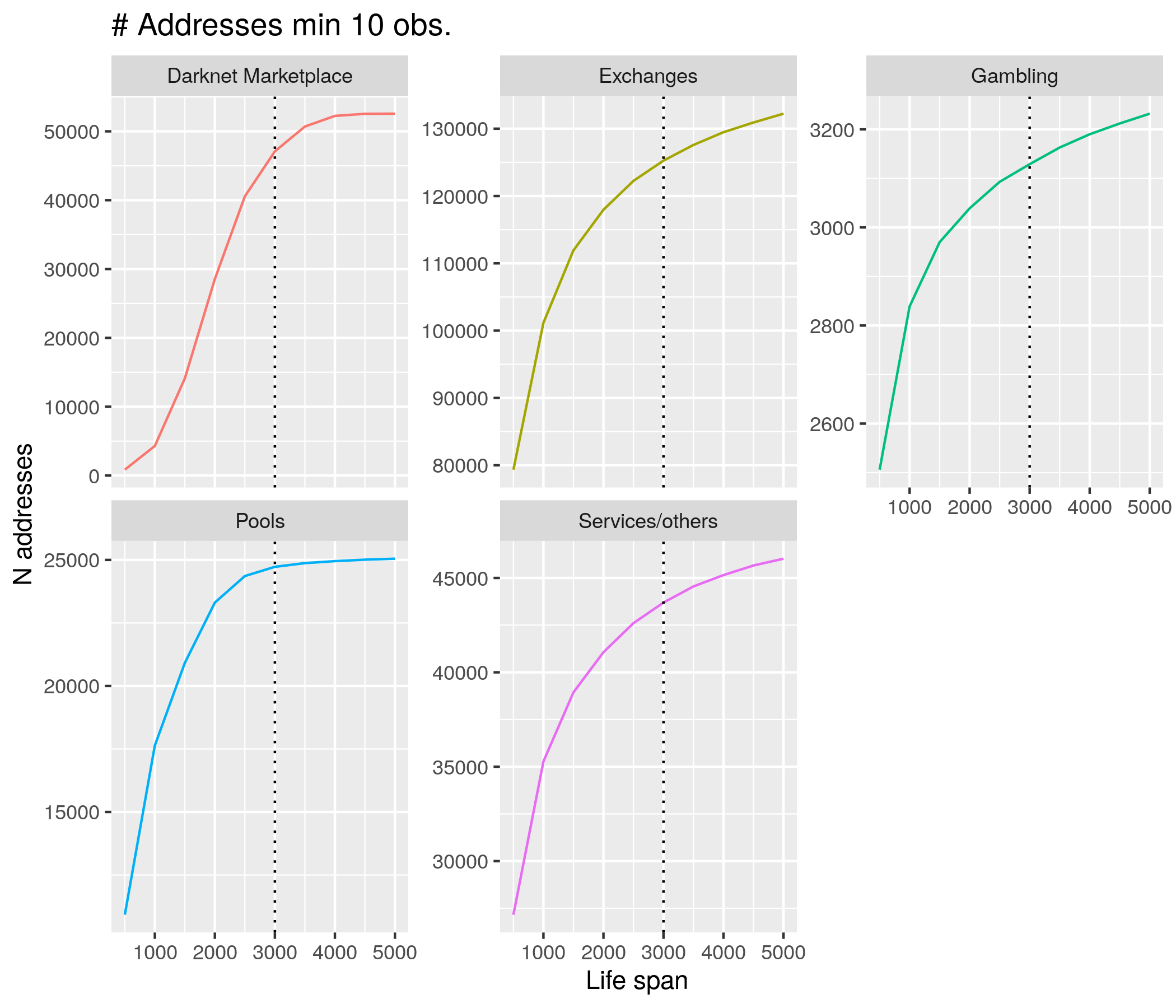}
    \caption{Number of addresses with at least 10 obs. by observed hours.}
    \label{fig:lifespan_10}
\end{figure}

Another important remark is the fact that one cannot really know if an address' life has ended, because there is nothing to prevent its use after a long time of inactivity. With that in mind, it was necessary to choose a time window to consider for prediction. The value was 3000 hours. Figure \ref{fig:lifespan_10} shows this as a reasonable value, for that are not a lot of addresses to gain after this span (to clarify, we first limit the observations to the first 3000 hours, and then filter the addresses with at least 10 or 20 account \final{transactions}). For addresses where a last transaction occurred before the end of this time frame, we know, since no other \final{transaction} was made, that the balances remains the same, so they were extended as constants until the end of the interval. This interval was then normalized, so that time extends between $[0,1]$.

\begin{figure}[H]
    \centering
    \includegraphics[scale = 0.5, center]{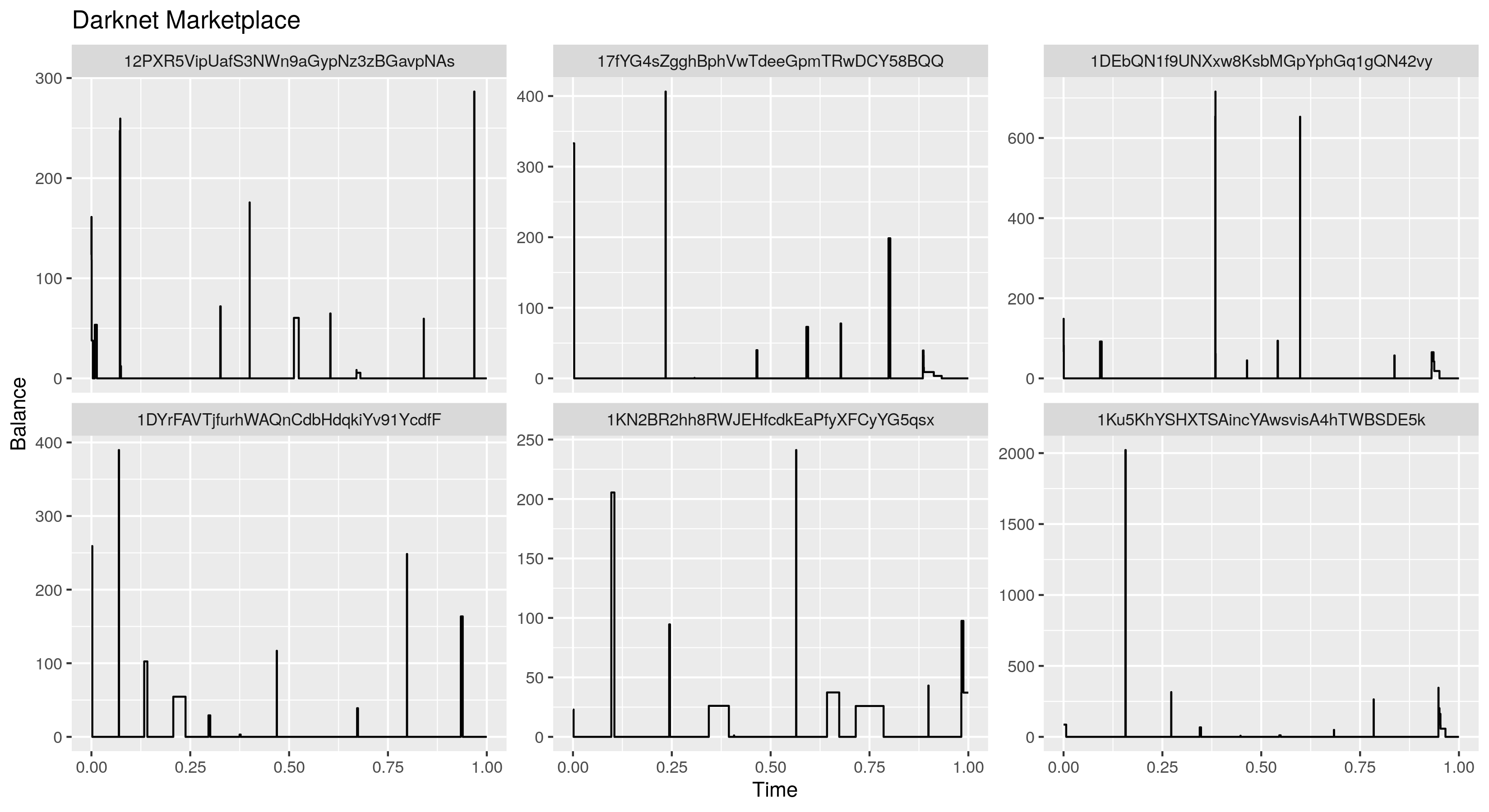}
    \caption{Original account balances for \final{six} darknet addresses, in dollars. \final{We will use these same addresses in the following figures.}}
    \label{fig:original_curves_darknet_20obs}
\end{figure}

The original data of the addresses' balances were very stiff; a lot of credits were immediately followed by debits of the same amount, which accounted for a very erratic behavior. Here, we show some examples for darknet addresses in Figure \ref{fig:original_curves_darknet_20obs}; \final{other categories can be viewed in the extended version of this paper \cite{fda_bitcoin_extended}}.

\begin{figure}[H]
    \centering
    \includegraphics[scale = 0.5, center]{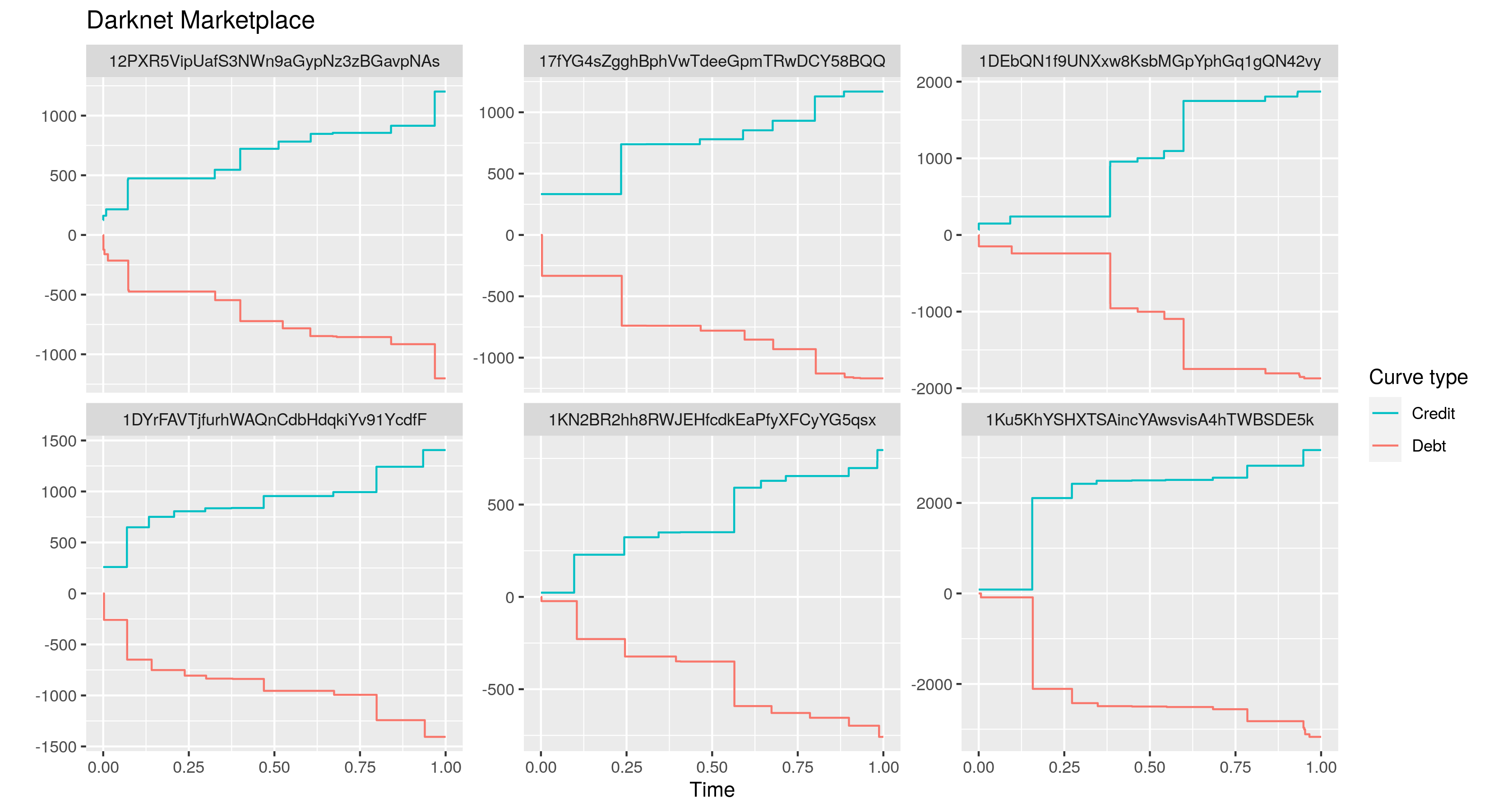}
    \caption{Accumulated sum of credits and debits for \final{six} darknet addresses, in dollars.}
    \label{fig:step_curves_darknet_20obs}
\end{figure}

The solution to this problem was to split the balances in order to form two different curves: one consisting of credits and another of debits. They were then integrated by calculating the accumulated sum (Figure \ref{fig:step_curves_darknet_20obs}). 

\subsection{Functional Data Modeling}\label{sec:func_data}

Even with a more regular behavior, the curves were still step functions. \final{We choose to smooth them because we want more regularity on our level measure. Even though there is no observational noise, in the sense that we know exactly when and how much was credited/debited at each address, since the underlying classification assumption is that each class follows a pattern of transactions, smoothing aids in capturing the more relevant trends in the amounts.} Through smoothing, one can represent the possibly infinite dimensional underlying function as an approximate finite dimensional basis expansion. To do so, we fit the data with a roughness penalty. 
%\yuri{In the following, we briefly present the methodology presented in \cite{RamsaySilverman2005book} in order to fix notation for the discussion of the results}.

\final{We consider the penalized functional regression framework over $L^2([0,1])$. The ingredients of this regression are the pre-defined basis of $L^2([0,1])$, denoted by $(\phi_k)_{k \in \mathbb{N}}$, the number $K \in \mathbb{N}$ of basis elements being considered in the approximation and the penalization, considered here as the $L^2$ norm of the $m$-th derivative of the approximation. The design matrix of regression, which is defined as the first $K$ elements of the pre-defined basis evaluated at the observation times, will be denoted by $\vec{\Phi}_{n \times K}$ and the penalization matrix $\vec{R}_{K \times K} = \int_0^1 D^m (\bphi(s))D^m (\bphi^T(s))ds$, where $\bphi = [\phi_1,\ldots,\phi_K]^T$.}

Because of the great variability in scale of the accumulated credits and debits across different addresses, the logarithm was also taken. 
%Therefore, in this case, the logarithm of accumulated credits (or debts) are $y_j$, and we use smoothing to estimate $x$. 
It is important to realize that it is possible to, regardless of the number of transactions, obtain an arbitrarily large number of observations through time for these curves - since it is known that, if no other transaction was made, the account value in bitcoins remains constant. In order to provide more data for better curve fitting, we used this fact to sample 501 evenly spaced points throughout the first 3000 hours additionally to the credit/debit times.

We choose the \textit{cubic} B-spline basis system, since it is the most common choice for polynomial functions. We penalized the second derivative, a natural measure of curvature ($m = 2$). One important aspect of using B-splines is choosing where to place the knots. \final{We chose to place at } almost every data point (including the original sampling points and the 501 evenly spaced points). It was necessary to remove points that were too close by (less than 3 minutes apart), because otherwise the steep nature of the data would make the estimation of derivatives, particularly the second derivatives, too large; which, in turn, would make the values of matrix $\vec{R}$ too high in magnitude, risking to overwhelm $\vec{\Phi}^T \vec{\Phi}$; $\vec{R}$ itself may not have full rank, so it may not be invertible; therefore, attempting to invert $(\vec{\Phi}^T \vec{\Phi} + \lambda\vec{R})$ will yield an error message or inaccurate results. 

The removal of close data points does not, however, imply loss of information, because the frequency is accounted for in the estimated Poisson rate curves, as will be explained in a couple of paragraphs. 
The smoothing parameter was chosen to be $\lambda = 0.001$. Although we did experiment with the Generalized Cross Validation (GCV) criterion %(citar alguém) 
to choose $\lambda$, we found the resulting curves to be under-smoothed, even after attempting to further discount the degrees of freedom. We wanted a reduced level of \final{roughness in the approximation}, \final{so we followed the heuristic approach} and our \final{visual} explorations \final{of fit} suggested $\lambda = 0.001$ to be a good choice. 
%This method is \yuri{heuristic}, \com{but, as stated by \cite{green1993nonparametric}, sometimes a \yuri{heuristic} approach is the most useful one.}

\begin{figure}[H]
    \centering
    \includegraphics[scale = 0.55, center]{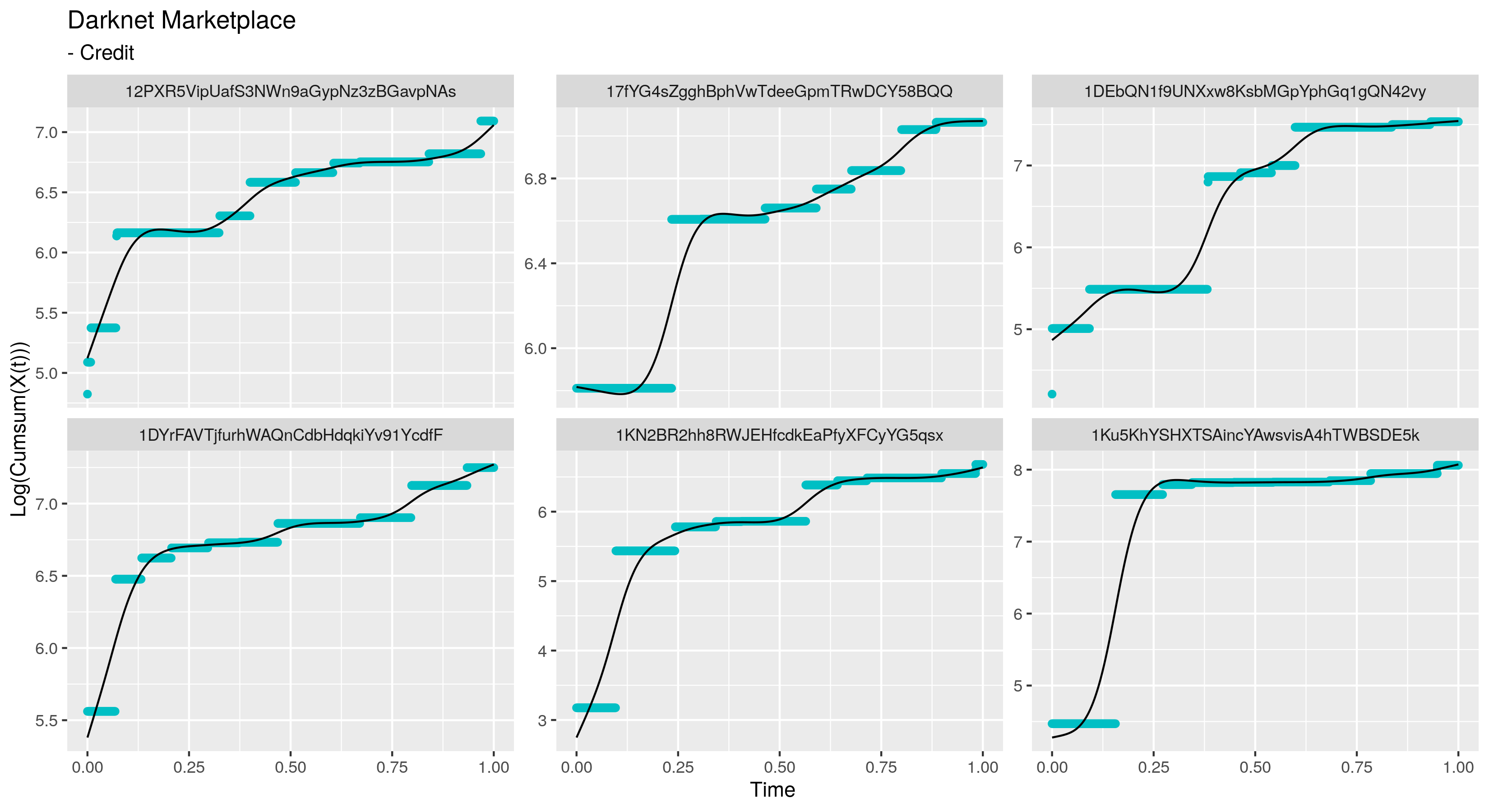}
    \caption{Smoothed log-accumulated sum of credits for \final{six} darknet addresses.}
    \label{fig:smooth_curves_darknet_20obs_credit}
\end{figure}

\final{Furthermore, using the same grid (evenly 501 points) as before, the functional principal component analysis was performed; it is convenient that these functions are regularly measured.}
%Therefore, we evaluate the smoothed function over the 501 evenly spaced points, for every address. 
Some examples for credit curves can be seen in Figure \ref{fig:smooth_curves_darknet_20obs_credit}.

\begin{figure}[H]
    \centering
    \includegraphics[scale = 0.55, center]{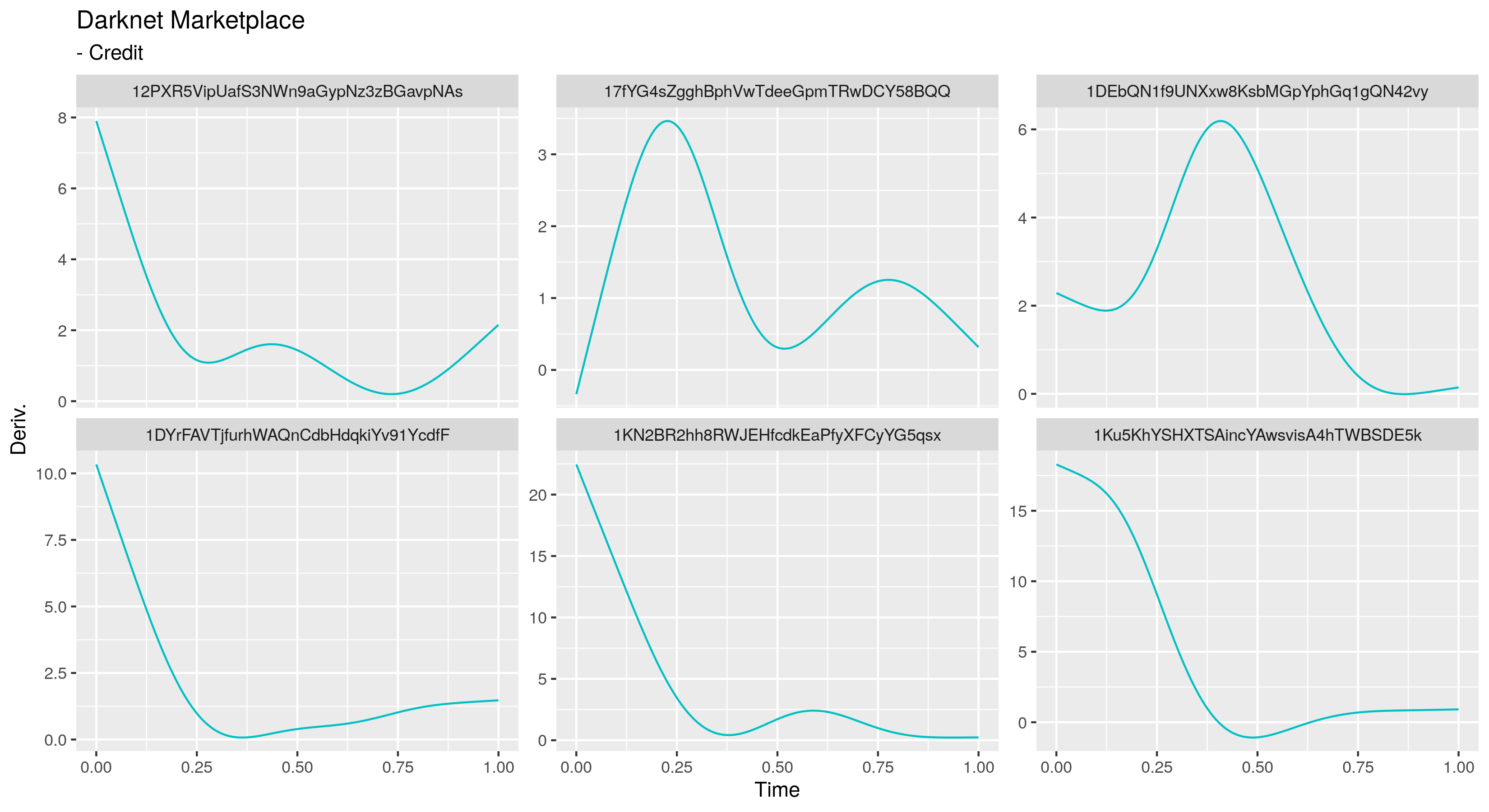}
    \caption{Derivatives of the smoothed log-accumulated sum of credits for \final{six} darknet addresses.}
    \label{fig:smooth_derivatives_darknet_20obs_credit}
\end{figure}

In order to better account for the variations in the curves, their first derivatives were also estimated. For a B-spline, the derivatives are calculated analytically after smoothing, but the details of such computations are not of particular interest to this paper. Since smoothness is a necessary hypothesis for the existence of derivatives, we must use the already smoothed curves $\hat{\vec{x}}(t) = \hat{\vec{c}}^T\bphi(t)$ rather than the step curves as our data. To estimate the curve's derivative of order $m$, the smoothing process is done by penalizing the derivative of order $m+2$ to ensure that the derivative itself will be smooth. We have then used an order $5$ B-spline system to re-smooth the curves applying \final{a penalization of the third derivative} ($5$ is the minimum order of a B-spline with a \final{square-}integrable third derivative). The derivatives were then evaluated at the same 501 equally spaced points. We chose $\lambda = 0.0001$ in the same fashion as the smoothing parameters for the curves themselves. The results for credits are displayed in Figure \ref{fig:smooth_derivatives_darknet_20obs_credit}. 

Furthermore, the curves can also be modeled by a Marked Point Process\final{, see for instance \cite{jacobsen2006point},} in which the credits and debits are viewed as arrivals. We simplify the hypothesis so that it becomes a inhomogeneous Poisson process. The assumption made for the functional model is that the rate is not constant over time, but a function of it instead. 

Denoting this inhomogeneous Poisson process by $(N(t))_{t \in [0,1]}$, its arrival times by $T_n$, $T_0 = 0$, and functional rate by $\mu:[0,1] \longrightarrow [0,+\infty)$, we have that, for instance, from Theorem 10.1 of \cite{thompson1988point},
$$T_1,\ldots,T_n \ | \ N(1) = n \sim f,$$
where $N(1)$ is the number of arrivals in $[0,1]$ and
\begin{align*}
&f(t_1,\ldots,t_n) = \frac{n!}{M^n} \prod_{i=1}^n \mu(t_i), \quad t_1 \leq t_2 \leq \cdots\leq t_n, \quad t_i \in [0,1], \quad M = \int_0^1 \mu(s)ds,
\end{align*}
where $t_1, \ldots t_n$ are the \final{observed} arrival times up to the $n$-th arrival. Therefore, if we observe the time of arrival of the $N(1) = n$ events that occurred in $[0,1]$, the likelihood becomes\footnote{Notice the trivial equality $\mathbb{P}(T_1 \leq t_1, \ldots, T_n \leq t_n, N(1) = n) = \mathbb{P}(T_1 \leq t_1, \ldots, T_n \leq t_n \ | \  N(1) = n) \times \mathbb{P}(N(1) = n).$}
\begin{align*}
L(t_1,\ldots,t_n \mid \mu) &= f(t_1,\ldots,t_n) \times \mathbb{P}(N(1) = n) \\
&= \frac{n!}{M^n} \prod_{i=1}^n \mu(t_i) \times e^{-M(1)} \frac{M^n}{n!}= e^{-M} \prod_{i=1}^n \mu(t_i),
\end{align*}
% \begin{equation}
% p(t_1,\ldots,t_N \mid \mu) = \exp\left[-\int_0^{t_N} \mu(t)dt\right] \prod_{i=1}^N \mu(t_i);
% \end{equation}
% see, for instance,  
Then, the log-likelihood function is given by
\begin{equation}
%    \begin{split}
        % L(t_1,\ldots,t_N | \mu) & = \prod_{n=1}^N \mu(t_n) \; \text{exp}\left [ - \int_{t_{n-1}}^{t_n} \mu(s) ds \right] \\
        % \implies 
        \text{log} [L(t_1,\ldots,t_n \mid \mu)] = \sum_{i=1}^n \text{log} \; \mu(t_i) - \int_0^1 \mu(s) ds.
%    \end{split}
\end{equation}

Following the \final{penalized functional regression framework}, we can write $\mu$ in a basis expansion. However, this function has a new constraint: $\mu(t) > 0$. To account for that, we take the exponential:
\begin{equation}
    \mu(t) = \text{exp} [\vec{c}^T \bphi(t)].
\end{equation}

% We may also penalize this fit with the penalty defined in Equation (\ref{pen_m}). Usually, since the simplified hypothesis is that $\mu$ is constant, the first derivative is penalized. By writing the penalty in vector form, as in Equation (\ref{argmin_basis}), the following problem must be numerically solved:
% \begin{equation}
% \label{eq:poisson_rate}
%     \begin{split}
%         & \argmax_{\vec{c}} \; \text{log} [L(t_1,\ldots,t_n \mid \mu)] - \lambda \; PEN_1(\text{log}[\mu]) \\
%         = & \argmax_{\vec{c}} \; \sum_{i=1}^n \vec{c}^T \bphi(t_i) - \int_\tau \text{exp}[\vec{c}^T \bphi(s)] ds - \lambda \times \vec{c}^T \vec{R} \vec{c}.
%     \end{split}
% \end{equation}
\final{Since the simplified hypothesis is that $\mu$ is constant, we penalize the log-likelihood maximization with the first derivative ($m=1$) of $\bphi$. Then, the following problem must be numerically solved:}
\final{\begin{equation}
\label{eq:poisson_rate}
    \begin{split}
    \argmax_{\vec{c}} \; \sum_{i=1}^n \vec{c}^T \bphi(t_i) - \int_\tau \text{exp}[\vec{c}^T \bphi(s)] ds - \lambda \times \vec{c}^T \vec{R} \vec{c}.
    \end{split}
\end{equation}}

\begin{figure}
    \centering
    \includegraphics[scale = 0.55, center]{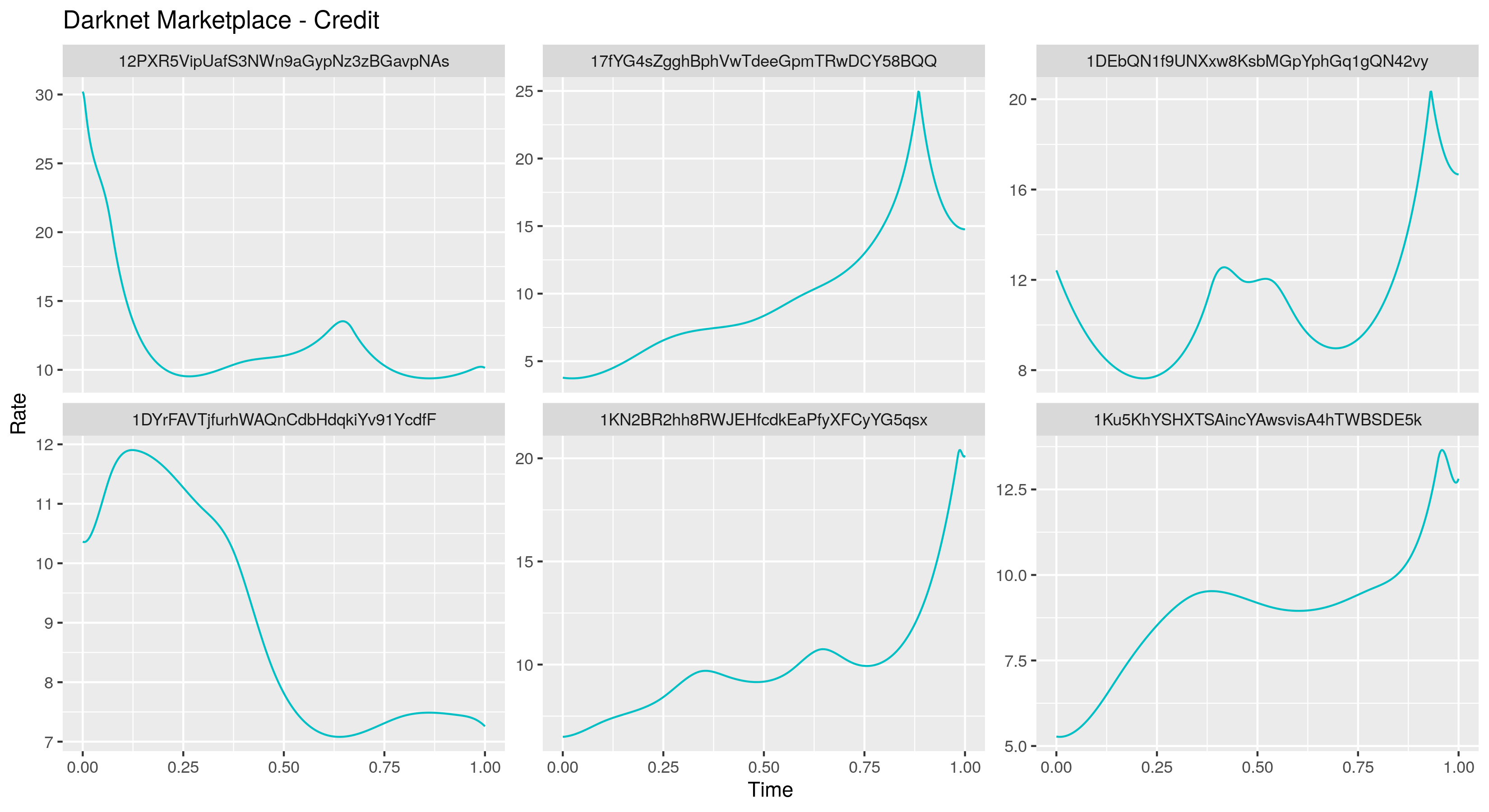}
    \caption{Poisson rate of credits for \final{six} darknet addresses.}
    \label{fig:poisson_rates_darknet_20obs_credit}
\end{figure}

We use the method described above, with a smoothing parameter of $\lambda=0.1$ and penalization on the first derivative \final{($m = 1$)}, for each credit and debit curve - recalling that the observations are considered for the first 3000 hours. \final{The parameter of $\lambda=0.1$ was also chosen heuristically.}
Results can be seen in Figure \ref{fig:poisson_rates_darknet_20obs_credit}. The resulting rate curves are less smooth than the other types of curves, which is natural given that the first derivative is used as penalization.

\section{Classification Models}
\label{sec:classification_models}

The prediction of the addresses' classes is a supervised classification problem, since the \final{classes} are known. Four different algorithms were used to compare models: Multinomial Logistic Regression, Gradient Boosting, Support-Vector Machine and Random Forest. Details of each algorithm can be found in \cite{friedman2001elements}.

There are essentially two types of features to build models in this case: scalar and functional. The scalar features are chosen according to the problem at hand, and have no higher wisdom behind their engineering than the domain knowledge of the person responsible for developing them. They are present in the vector model described below. Functional features, on the other hand, are less arbitrary in the sense that they are representations of the function being analyzed. This is a result of the basis regression minimization problem expressed in Section \ref{sec:func_data}. The domain specific choices, in this case, are the basis system, the number of basis functions and the smoothing parameter.

\subsection{Vector Model}
The simplest model attempted is a vector model with scalar features built from credits and debts transactions. There is no functional treatment in this approach. The features considered here are: credit/debt count; credit/debt sum; credit/debt minimum; credit/debt maximum; credit/debt median; credit/debt first quantile; credit/debt third quantile; difference of third and first quantile values for credits/debts; credit/debt interval time (time elapsed between first and last credits/debts).
The features regarding the difference of quantiles were not used with the multinomial logit model in order to avoid multicollinearity.

Additionally, constant Poisson rates (number of credits divided by interval time/ number of debts divided by interval time) were added in order to compare the gains of a functional Poisson rate.

\subsection{Functional Model}
We choose to adapt the multivariate classification algorithms to the functional context for one main reason: we want to be able to compare and combine models using features accounting for different curves and the vector approach.

As previously stated, the functional features are representation of the functions at hand; more specifically, they are the coefficients of the basis expansion of each curve. For this work, we choose the functional principal components (FPCs) basis, as it usually allows for the representation of the majority of the function's variance with few basis functions, thus preventing overfitting. In this case, the coefficients are the FPCs. We estimate them for the training set as described in Section 8.4.2 of \cite{RamsaySilverman2005book}, for each curve type (accumulated balances, first derivative and Poisson rates for each credit and debt). This classical method requires the data to be sampled at the same timestamps. Then, the equally sampled data must be written on a previously defined basis: similarly to the smoothing procedure described in Section \ref{sec:func_data}, we use B-splines to re-smooth the data, with penalization on the second derivative and knots at equally spaced times. Different re-smoothing parameters are tested in terms of classification performance; more details are given in Section \ref{sec:results}.

Notice that the coefficients of the re-smoothing procedure are different from the first smoothing procedure described in Section \ref{sec:empirical_strategy}, in which each function is smoothed individually, with different timestamps, in a pre-defined basis. However, smoothing data on the FPC basis cannot be done individually, because the FPC basis is an empirical basis; that is, smoothing on this basis is dependent on the set of observed curves. Since we are building a prediction model, it is important to differentiate how the principal components are computed for the training and test sets: the FPC basis functions are estimated on the training set, and then the test data set is smoothed over it. The validation data set is also re-smoothed in the same fashion. 

However, to obtain the FPCs for the test set, we need to find the coefficients of each curve corresponding to the expansion on the eigenbasis estimated on the training set. 
Let $\vec{\Xi}$ be the $n\times L$ matrix containing the values $\hat{\xi}_i(t_j)$, where $\hat{\xi}_i$ is the approximation of the $i$th eigenfunction of the covariance operator, $\hat{\xi}_i(t) = \vec{b}_i^T\bphi(t)$, and we are considering the first $L$ eigenfunctions. Hence
\begin{equation}
    \vec{\Xi} = \vec{\Phi}\vec{B},
\end{equation}
where $\vec{B}$ is the $K \times L$ matrix whose columns are the coefficients of each eigenfunction in the given basis $\phi$ and $\vec{\Phi}$ as defined in Section \ref{sec:func_data}. If $\vec{Z} = [Z_1,\ldots,Z_K]^T$ is the vector of FPCs, we can estimate it as in Section \ref{sec:func_data}:
\begin{equation}
    \hat{\vec{Z}} = \left ( \vec{\Xi}^T\vec{\Xi} + \lambda \vec{R}^{*} \right )^{-1}\vec{\Xi}^{T} \vec{y_{cen}},
\end{equation}
with $\vec{y_{cen}}$ being the centered observation $\vec{y} - \hat{\bmu}$, where $\hat{\bmu} = [\hat{\mu}(t_1),\ldots,\hat{\mu}(t_n)]$ and $\hat{\mu}$ is the mean of the smoothed curves in the training set at every $t_j$. Moreover, $\vec{R}^{*}$ is the penalization matrix of $\vec{\Xi}$ \final{and it is given by $\vec{R}^{*} = \vec{B}^T \vec{R} \vec{B}$, where $\hat{\bxi} = [\hat{\xi}_1,\ldots,\hat{\xi}_L]^T$.}
% \begin{equation}
%     \begin{split}
%         \vec{R}^{*} & = \left [ \int_{\tau} D^{m}(\hat{\bxi}(s))D^{m}(\hat{\bxi}^T(s))ds \right ] \\
%         &= \left [ \int_{\tau} D^{m}(\vec{B}^T\bphi(s))D^{m}(\bphi^T(s)\vec{B})ds \right ] \\
%          & = \vec{B}^T \left [ \int_{\tau} D^{m}(\bphi(s))D^{m}(\bphi^T(s))ds \right ] \vec{B}  = \vec{B}^T \vec{R} \vec{B},
%          %& = \vec{b}^T \left [ \int_{\tau} D^{m}(\bphi(s))D^{m}(\bphi^T(s))ds \right ] \vec{b} \\
%          %& = \vec{b}^T \vec{R} \vec{b}.
%     \end{split}
% \end{equation}

\section{Results}
\label{sec:results}

Models such as the Multinomial Logistic Regression, Support-Vector Machine, Gradient Boosting and Random Forest accept multiple functional objects and also vector objects, which means that these features can be combined to build more accurate classifiers.\footnote{\final{Code and data for replicating this paper can be found at \url{https://github.com/brendaprallon/FDA-bitcoin-paper}.}} In the following section, the results for different combinations will be presented. The combinations are: 

\begin{itemize}
    \item Vector Model (only);
    \item Functional Model (with all types of curves);
    \item Vector Model combined with Functional Model;
    \item Vector model with constant Poisson rates;
    \item Vector model with functional Poisson rates;
    \item Vector model with functional derivatives and functional Poisson rates.
\end{itemize} 

Regarding the prediction of the classes, undersampling was employed with the purpose of getting a balanced dataset (see Table \ref{n_obs_cat_orig}): for a minimum of 10 \final{transactions}, we select 3129 samples of each \final{class} (the minimum number of addresses in one category - gambling); and for 20, 1741 samples (also the minimum of gambling addresses).
20\% of the samples were set apart for testing the best model, and the best model was chosen by 5-fold cross-validation in the training set (80\% of samples). 

Multiple values for different parameters were tested:  re-smoothing parameters of $10^{-10}$ and $10^{-1}$; and, for the number of principal components to use to describe each functional attribute, 1, 3, 5, and 7 FPCs were used; however, they were tested for all the functions at a time, that is, one model considered 1 principal component for credit curves, 1 for debit curves, 1 for credit derivatives, 1 for debit derivatives, 1 for credit rates, 1 for debit rates; another model considered 3 principal components for credit curves, 3 for debit curves, 3 for credit derivatives... and so on.

\begin{figure}[H]
    \centering
    \includegraphics[scale = 0.5, center]{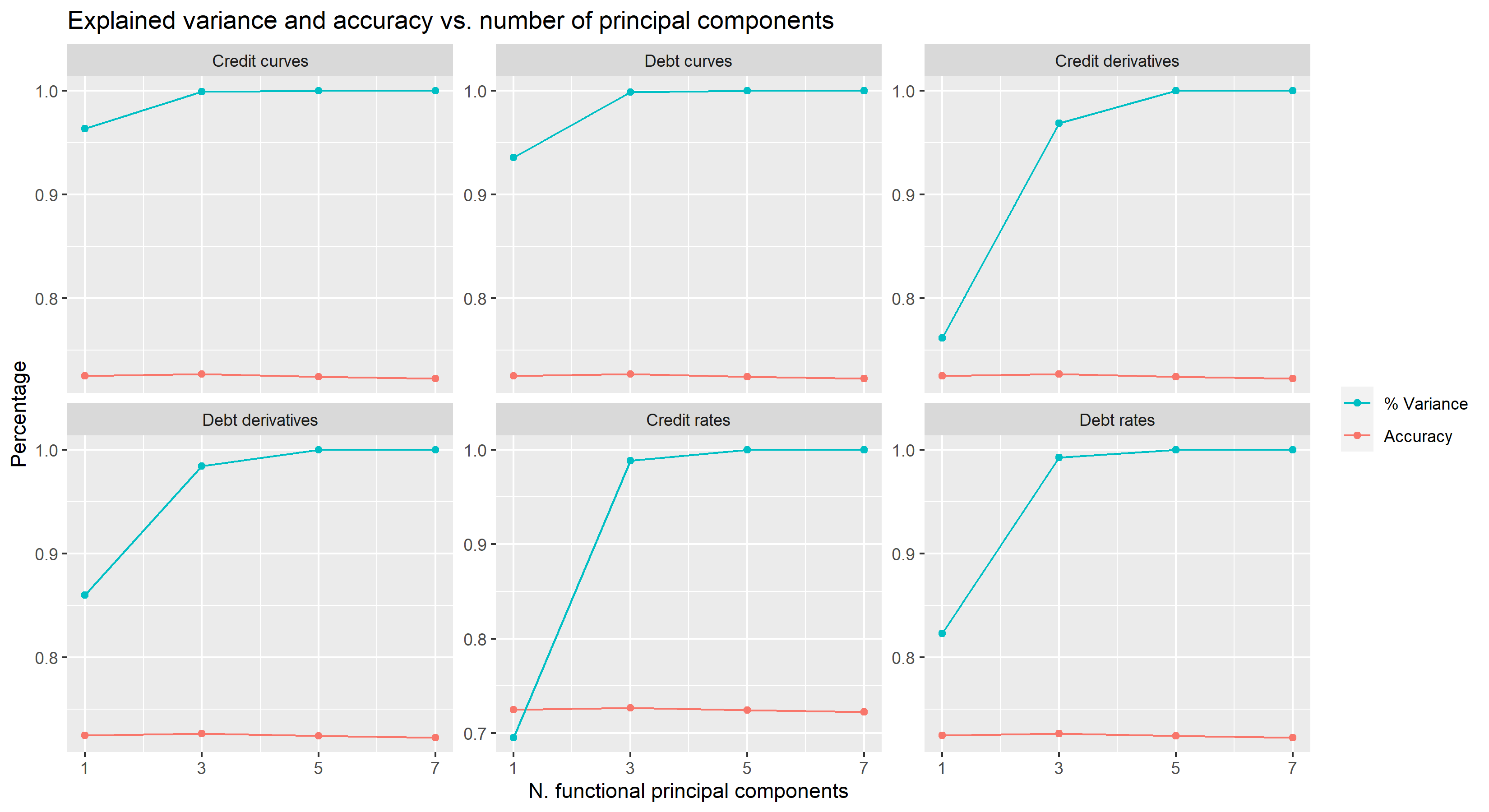}
    \caption{Accuracy and proportion of explained accumulated variance for the winning model with different numbers of principal components, min. 20 obs.}
    \label{fig:acc_vs_variance_20obs}
\end{figure}

\final{Because we have performed undersampling to ensure the model was balanced - that is, we have the same number of observations for each class - evaluation of the models was done by computing their accuracy.} In the validation process, overall, the random forest classification algorithm performed better, regardless of the number of principal components, re-smoothing parameter, or minimum \final{number of transactions (10 or 20)}. Few exceptions were the vector model (only) or functional model, which for some combination of number of principal components and smoothing parameters were outperformed by the Gradient Boosting algorithm with both functional and vector features. The re-smoothing parameter did not seem to make a significant difference on the result -- this makes sense, considering the curves were already smooth and the re-smoothing process is just part of the FPC estimation; neither did the number of FPCs (in Figure \ref{fig:acc_vs_variance_20obs}, we can see that the number of principal components does not greatly affect the accuracy). Instead, what seems to be the most relevant factor for better accuracy are the combined functional and scalar features. The vector model combined with functional model, vector model with functional Poisson rates, and vector model with functional derivatives and functional Poisson rates consistently outperform the vector model (only), functional model (with all types of curves), and vector model with constant Poisson rates. 

To properly access the accuracy of the estimation, the models were run with the best parameters for each algorithm using a test set. We report the results for the random forest algorithm, given that it was consistently better during validation. For a minimum of 10 observations, the best model was the vector model with functional derivatives and functional Poisson rates, re-smoothing parameter of $10^{-10}$ and 5 FPCs; for a minimum of 20 observations, this was the vector model combined with functional model, re-smoothing parameter of $0.1$ and only 1 FPC.

Results for a minimum of 10 and 20 observations can be seen in Tables \ref{rf_10obs_acc} and \ref{rf_20obs_acc}, respectively. Contrary to what was expected, there were no significant improvements when further restricting the sample to a minimum of 20 observations: the accuracy of the best models are around $0.72$. However, as was observed during validation, combining scalar and functional features has the potential to improve accuracy between $3-5\%$, and it seems that functional features of derivatives and rates contribute more than the functional features of the curves themselves. Furthermore, it is worth noticing the functional model achieves similar results to the vector model - outperforming it by $3\%$ in the case of a minimum of 10 observations, and being short of no more than $2\%$ when using the 20 observations threshold. This shows that in a setting where the feature engineering process is too complex or where specific domain knowledge is scarce, representing the data with functional principal components could be a very good alternative, given that it is a straightforward method.

It does not appear that increasing the number of FPCs would aid in capturing more information. Figure \ref{fig:acc_vs_variance_20obs} show that 3 components explain the variance almost entirely, and that the effect of the number of components on accuracy is very limited. However, even though it is evident that the scalar features are important for classification, the information contained on the empirical basis of the curves mimic them quite well, even with few components.

\begin{figure}
    \centering
    \includegraphics[scale = 0.4, center]{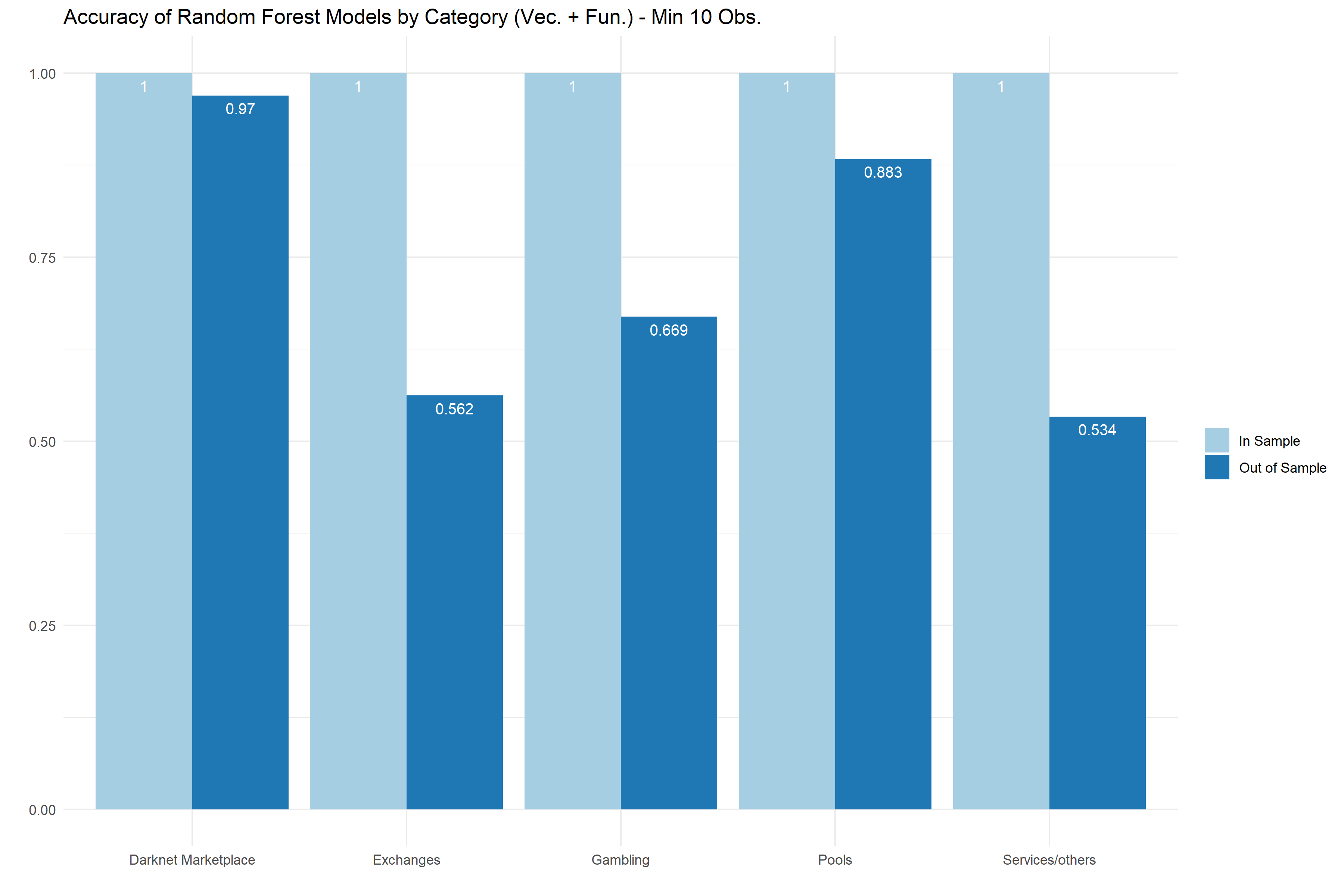}
    \caption{Accuracy by category of the combining model using all scalar and functional features, min. 10 obs.}
    \label{fig:acc_cat_rf_10obs}
\end{figure}

\begin{figure}
    \centering
    \includegraphics[scale = 0.4, center]{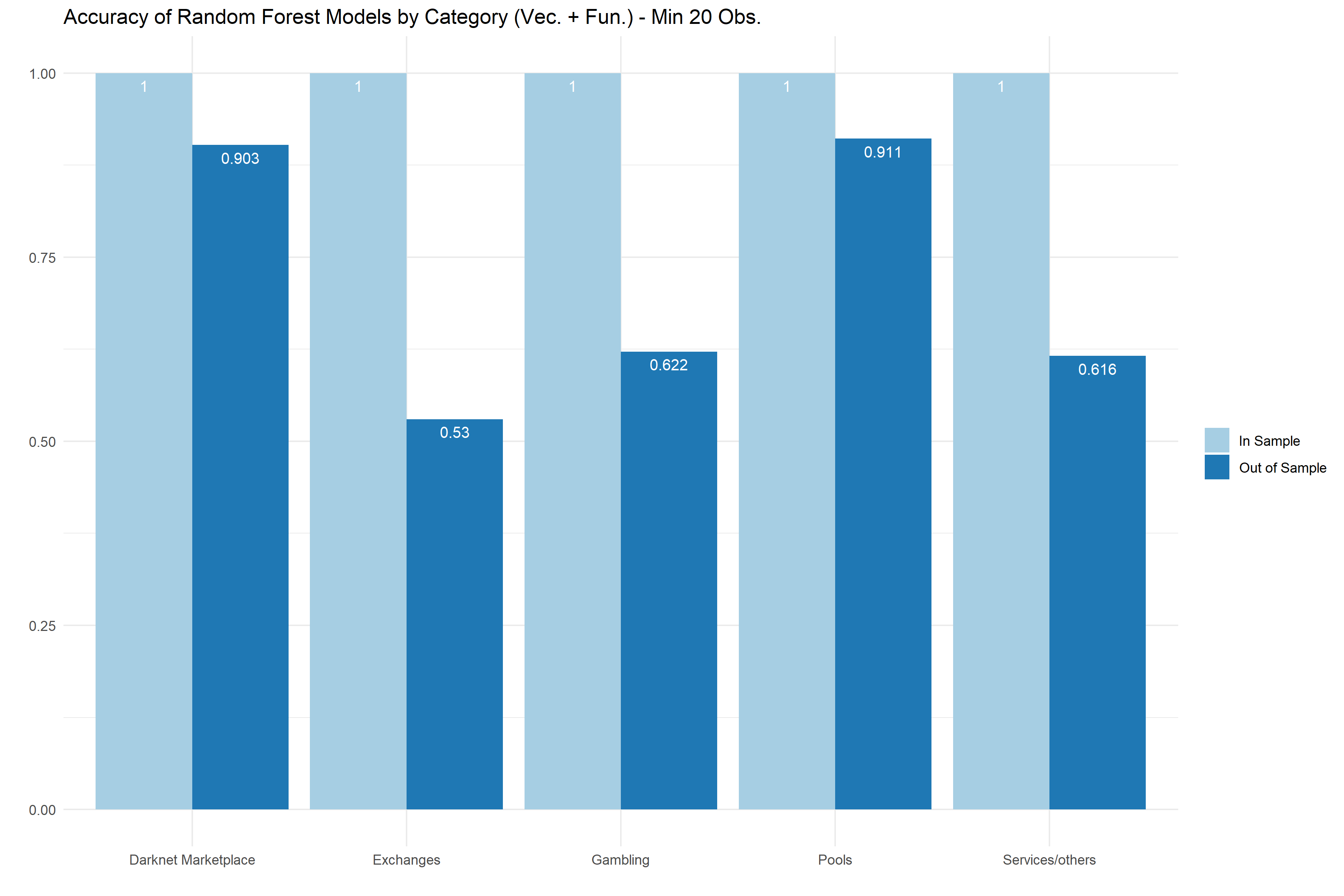}
    \caption{Accuracy by category of the combining model using all scalar and functional features, min. 20 obs.}
    \label{fig:acc_cat_rf_20obs}
\end{figure}

The fact that the darknet and pool addresses are consistently better classified (Figures \ref{fig:acc_cat_rf_10obs} and \ref{fig:acc_cat_rf_20obs}) could mean that the sampling issue described in Section \ref{sec:sampling_issue} is still present. The behavior of the addresses might change along with the bitcoin price, or the behavior is more consistent along the same entity, and categories with fewer entities would be easier to classify. This is corroborated by the low accuracy of the exchange category, which has 50 entities. 

\begin{figure}
    \centering
    \includegraphics[scale = 0.5, center]{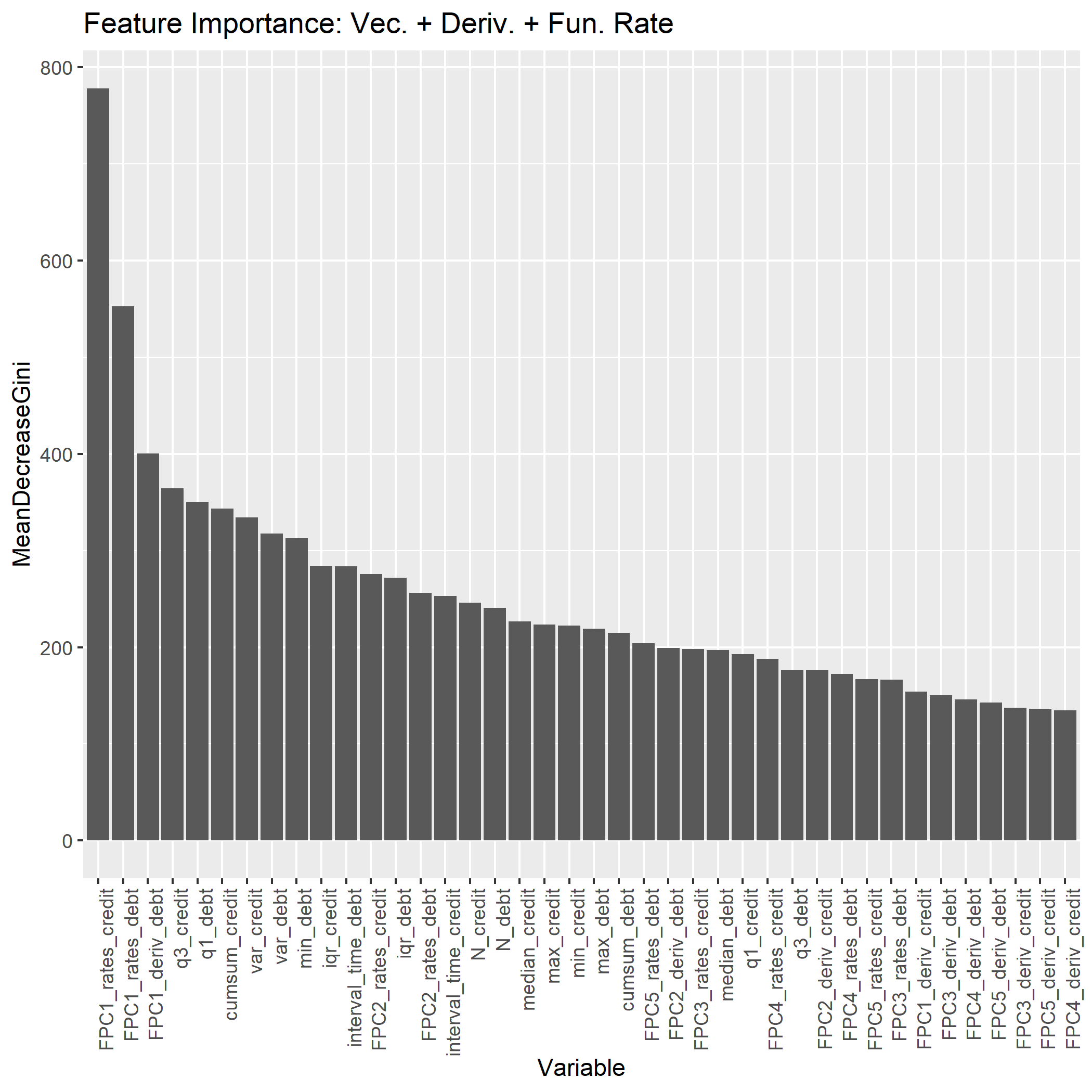}
    \caption{Feature importance of the combining model using scalar, functional rates and derivatives, min. 10 obs.}
    \label{fig:feature_importance_model_vec_fun_10obs}
    \includegraphics[scale = 0.5, center]{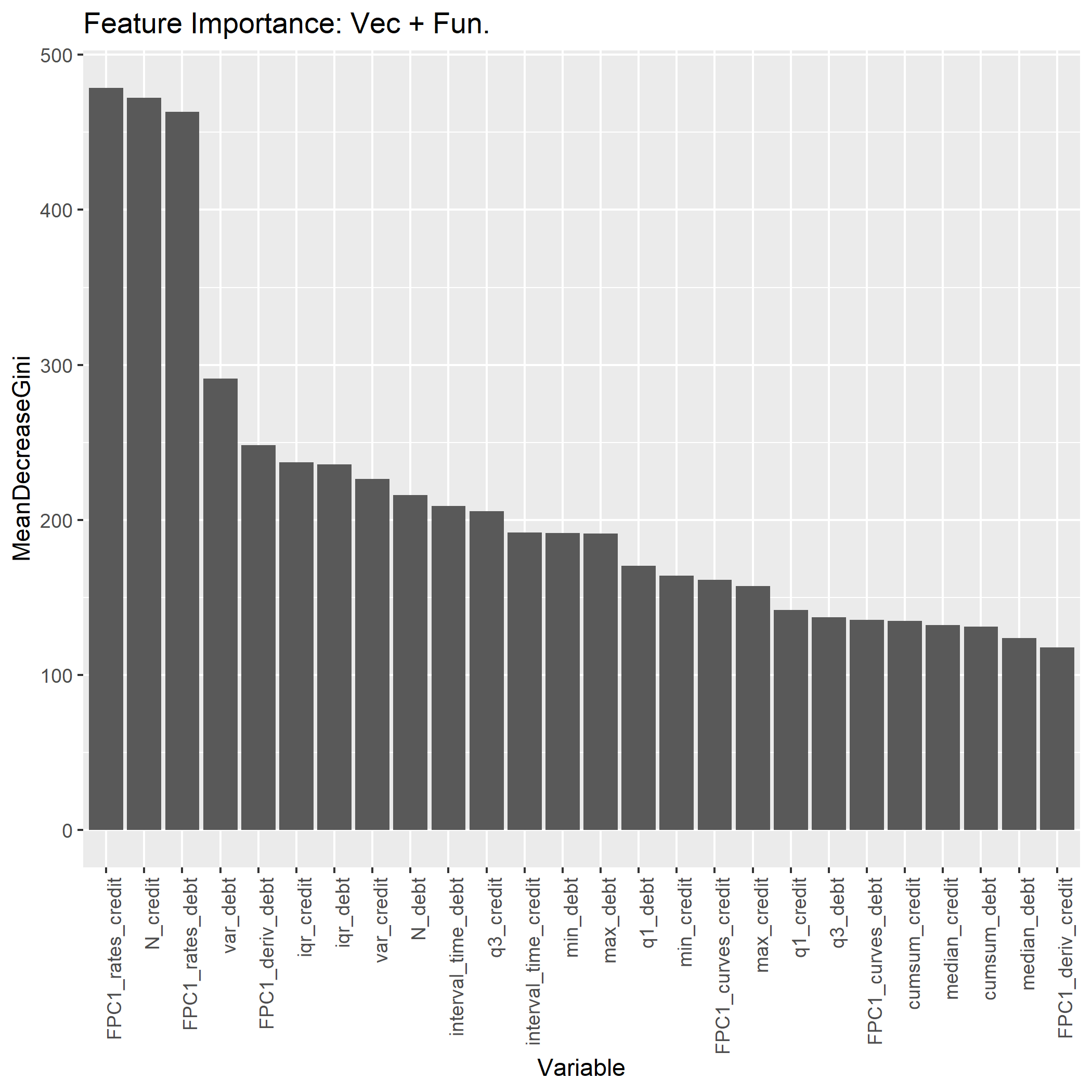}
    \caption{Feature importance of the combining model using all scalar and functional features, min. 20 obs.}
    \label{fig:feature_importance_model_vec_fun_20obs}
\end{figure}

The feature importance plot of the random forest model, for a minimum of 10 observations (Figure \ref{fig:feature_importance_model_vec_fun_10obs}), points to the first FPC of the credit rates curves as being the most relevant. It is followed, with a significant difference, by the first FPC of the debit rates curves, and then the first FPC of the debit derivative; the other variables are much closer in terms of aiding prediction. For a minimum of 20 observations (Figure \ref{fig:feature_importance_model_vec_fun_20obs}), the first FPC of credit and debit rates curves continue to be among the three more important, but the number of credits gain more importance. The relative importance of the three most important features to the followings is also large. It is clear that functional frequency measures are the most helpful in prediction, specially when considering addresses with fewer \final{transactions}. 

\begin{table}[ht]
\begin{adjustbox}{width=1.0\textwidth,center}
\centering
\begin{tabular}{rrrrrrr}
  \hline
 & Vec. & Vec. + Const. Rate & Vec. + Fun. Rate & Vec. + Fun. & Fun. & Vec. + Deriv. + Fun. Rate \\ 
  \hline
In Sample & 1.000 & 1.000 & 1.000 & 1.000 & 1.000 & 1.000 \\ 
  Out of Sample & 0.671 & 0.680 & 0.702 & 0.724 & 0.702 & 0.720 \\ 
   \hline
\end{tabular}
\end{adjustbox}
\caption{Accuracy of Random Forest models - Min 10 Obs.} 
\label{rf_10obs_acc}
\end{table}

% latex table generated in R 3.6.3 by xtable 1.8-4 package
% Thu May 20 09:52:26 2021
\begin{table}[ht]
\centering
\begin{adjustbox}{width=1.0\textwidth,center}
\begin{tabular}{rrrrrrr}
  \hline
 & Vec. & Vec. + Const. Rate & Vec. + Fun. Rate & Vec. + Fun. & Fun. & Vec. + Deriv. + Fun. Rate \\ 
  \hline
In Sample & 1.000 & 1.000 & 1.000 & 1.000 & 1.000 & 1.000 \\ 
  Out of Sample & 0.688 & 0.691 & 0.713 & 0.716 & 0.673 & 0.723 \\ 
   \hline
\end{tabular}
\end{adjustbox}
\caption{Accuracy of Random Forest models - Min 20 Obs.} 
\label{rf_20obs_acc}
\end{table}

%\subsubsection{Other Algorithms}
%\input{tables/other_algorithms_acc_10obs}

%\subsubsection{Gradient Boosting}

%\input{tables/gbm_10obs_acc}

%\input{tables/gbm_10obs_acc_cat}

%\subsubsection{Support Vector Machine}

%\input{tables/svm_10obs_acc}

%\input{tables/svm_10obs_acc_cat}

%\subsubsection{Multinomial Logit}

%\input{tables/multi_10obs_acc}

%\input{tables/multi_10obs_acc_cat}

% \subsection{Min. 20 observations}

% \subsubsection{Random Forest}

%\input{tables/rf_20obs_acc_cat}

\subsection{Functional Principal Components}

\begin{figure}
\centering
   \begin{subfigure}{0.45\linewidth}
   \centering
   \includegraphics[width=\linewidth]{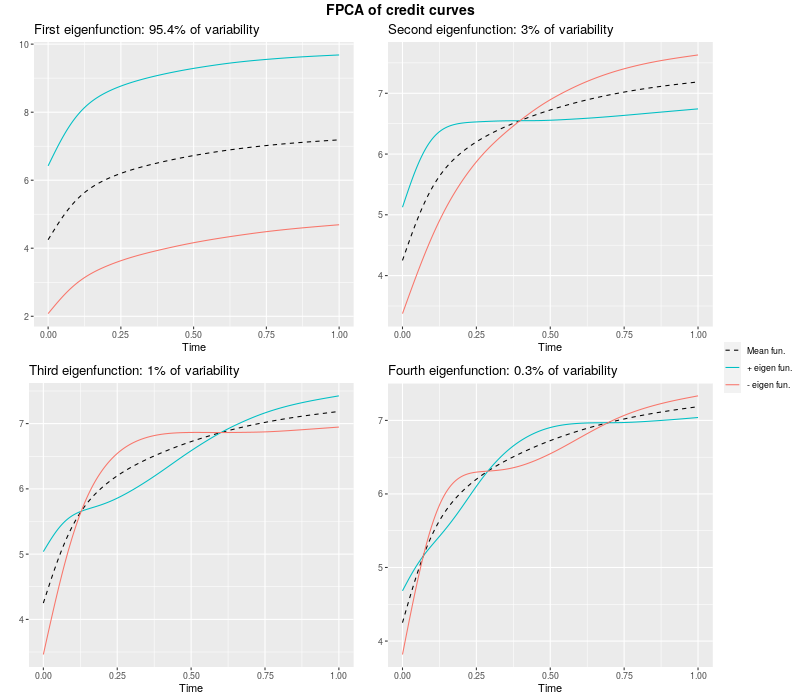}
   \caption{}
   \label{fig:fpca_curves_credit_20obs} 
\end{subfigure}
\hfill
\begin{subfigure}{0.45\linewidth}
   \centering
   \includegraphics[width=\linewidth]{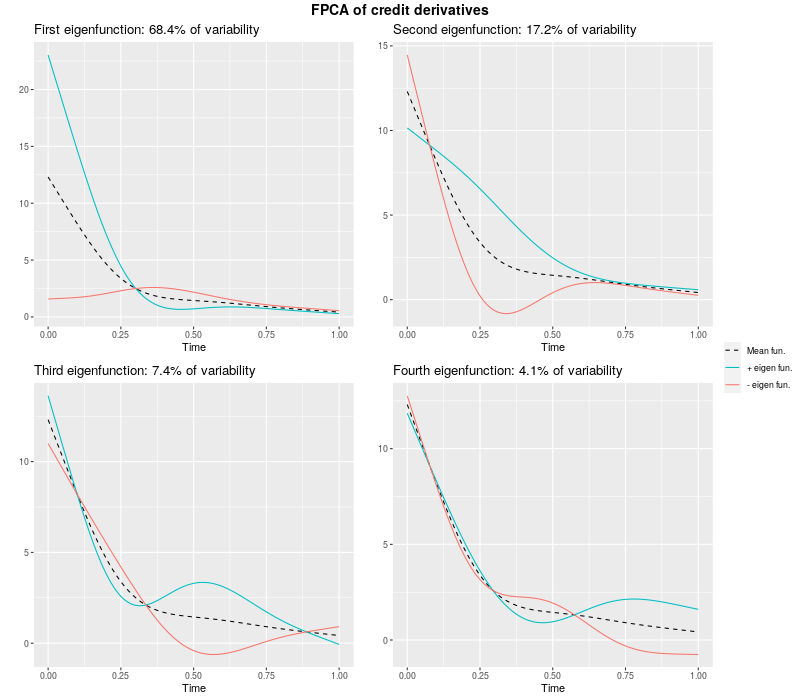}
   \caption{}
   \label{fig:fpca_derivatives_credit_20obs}
\end{subfigure}
\\[\baselineskip]
\begin{subfigure}[H]{0.45\linewidth}
   \centering
   \includegraphics[width=\linewidth]{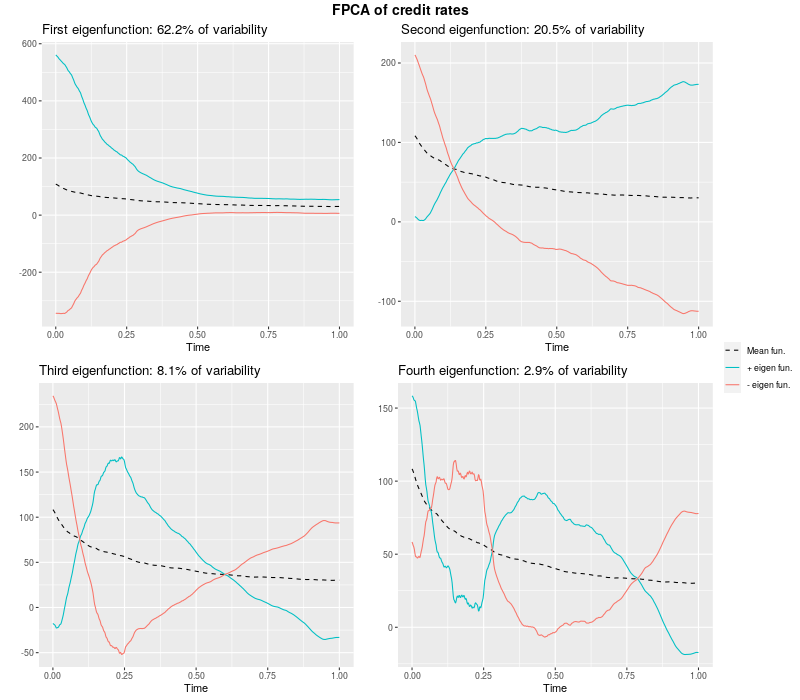}
   \caption{}
   \label{fig:fpca_rates_credit_20obs}
\end{subfigure}
\centering
\caption{(a) FPCA of credit curves, min. 20 obs. (b) FPCA of credit derivatives, min. 20 obs. (c) FPCA of credit rates, min. 20 obs.}
\label{fig:fpca_plots}
\end{figure}

Figure \ref{fig:fpca_plots} shows resulting plots of adding and subtracting eigenfunctions (weighted by their standard deviations) around the mean curve. The plots are made with the balanced sample with a minimum of 20 observations; the equivalent plots for a minimum of 10 observations are very similar. The modes of variation of credits and debits are very alike for the log curves and derivatives, but vary a bit for the rates, even though they preserve the general behavior. Rates are also an exception when it comes to the smoothness of eigenfunctions; they are significantly rougher than their counterparts. This is expected since the rate curves are themselves rougher because penalization is on the first derivative, and no smoothness constraints were imposed for the resulting eigenfunctions.

Regarding the curves, the first mode of variation represents differences in level: this accounts for more than 95\% of variance of accumulated credits, and over 90\% of accumulated debits. The next eigenfunctions represent shifts around the mean: curves that start with more (less) volume than the mean, and, at some point in time, invert this relation. For derivatives, the most important mode of variation captures the behavior of curves that are faster (slower) than the mean at the beginning, shift their trend after about one quarter of the addresses life and than stabilize around the mean. Finally, the variability of rate curves is maximized by weighting heavily the initial \final{instants} of the addresses life. This makes sense since many addresses have \final{transactions} up until before 3000 hours, which concentrates the frequencies in the beginning. 

\section{Conclusion}
\label{sec:conclusion}

FDA methods were employed to properly represent the \final{addresses' balances} as functions; some treatment was required, such as removing addresses with few observations, and smoothing was applied to obtain the log curves, the derivatives and the Poisson rates. The functional principal component basis was used. At last, different models were estimated using functional, scalar and a combination of both variables. The functional variables were the FPCs given by smoothing procedures.

It has been shown that, for the available data, the random forest algorithm yields better results in general than the other tested algorithms (multinomial logit, SVM and gradient boosting). The best results are given by models combining scalar and functional variables: improvements are around 5\% for a threshold of 10 observations. This points out that variance might not be the best way to separate the data in this case. However, the fact that the accuracy of the functional and vector model are very close highlights an important point: the principal components basis is capable of providing automatic features that are enough to replace domain-specific features. This can be specially useful in contexts where domain-specific knowledge is not available, or typical feature engineering is too costly. An additional highlight is that the model is capable of predicting classes without network information: for example, it is possible to predict addresses of new entities, which have possibly never interacted with any other address on the training set. \final{This is not the common case in the literature.} \final{and is our main contribution.}

Furthermore, the darknet category has the best accuracy, while the exchange category has the worst. Given the distribution of their addresses over time, this could mean that there is still an implicit timestamp bias. The number of entities associated with each category may also contribute to the \final{difference in} accuracy across groups. One possible solution is to gather the addresses per entity and classify them instead, but more samples would be necessary. Another is to explore notions of depth for classification of more heterogeneous data. These are interesting proposals for future work.

\section*{Acknowledgments}
\final{YFS and BP were partly supported by Silicon Valley Community Foundation and by the Coordenação de Aperfeiçoamento de Pessoal de Nível Superior – Brasil (CAPES) – Finance Code 001. The research by Manuel Febrero--Bande and Wenceslao González--Manteiga has been partially supported by the Spanish Grant PID2020-116587GB-I00 funded by MCIN/AEI/10.13039/501100011033  through the European Regional Development Funds (ERDF).
We thank the Associate Editor and two anonymous referees for their insightful comments, which improved the
quality and clarity of our paper substantially.}

 \appendix
 \section*{Appendix}
 \renewcommand{\thesection}{\Alph{section}}

 \section{Graphical Appendix}
 \label{sec:app}

 \subsection{Plots of addresses of other categories}
\label{plots_other_cats}

\subsubsection{Balances}

\begin{figure}[H]
    \centering
    \includegraphics[scale = 0.55, center]{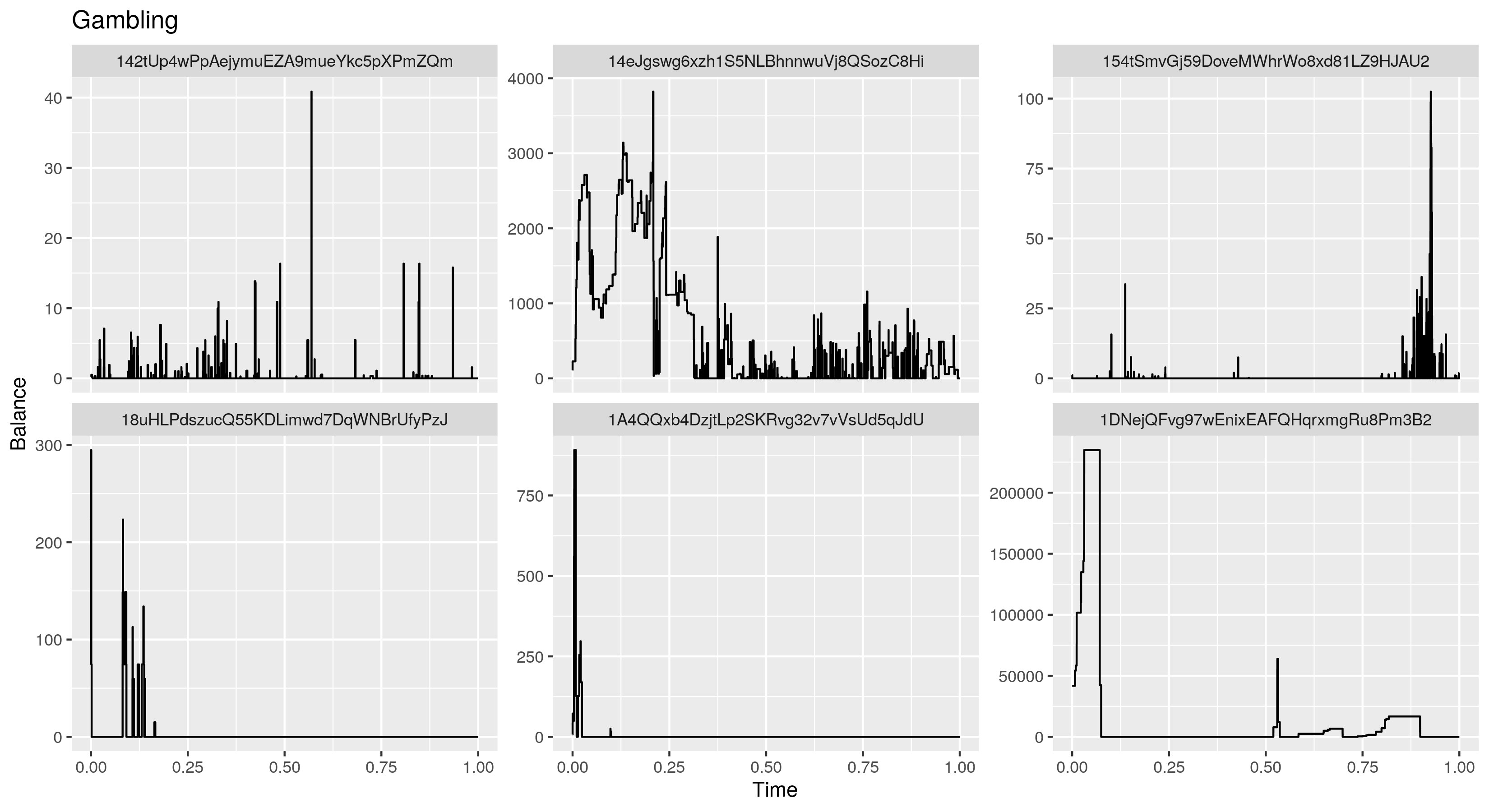}
    \caption{Original account balances for \final{six} gambling addresses, in dollars.}
    \label{fig:original_curves_gambling_20obs}
\end{figure}

\begin{figure}[H]
    \centering
    \includegraphics[scale = 0.55, center]{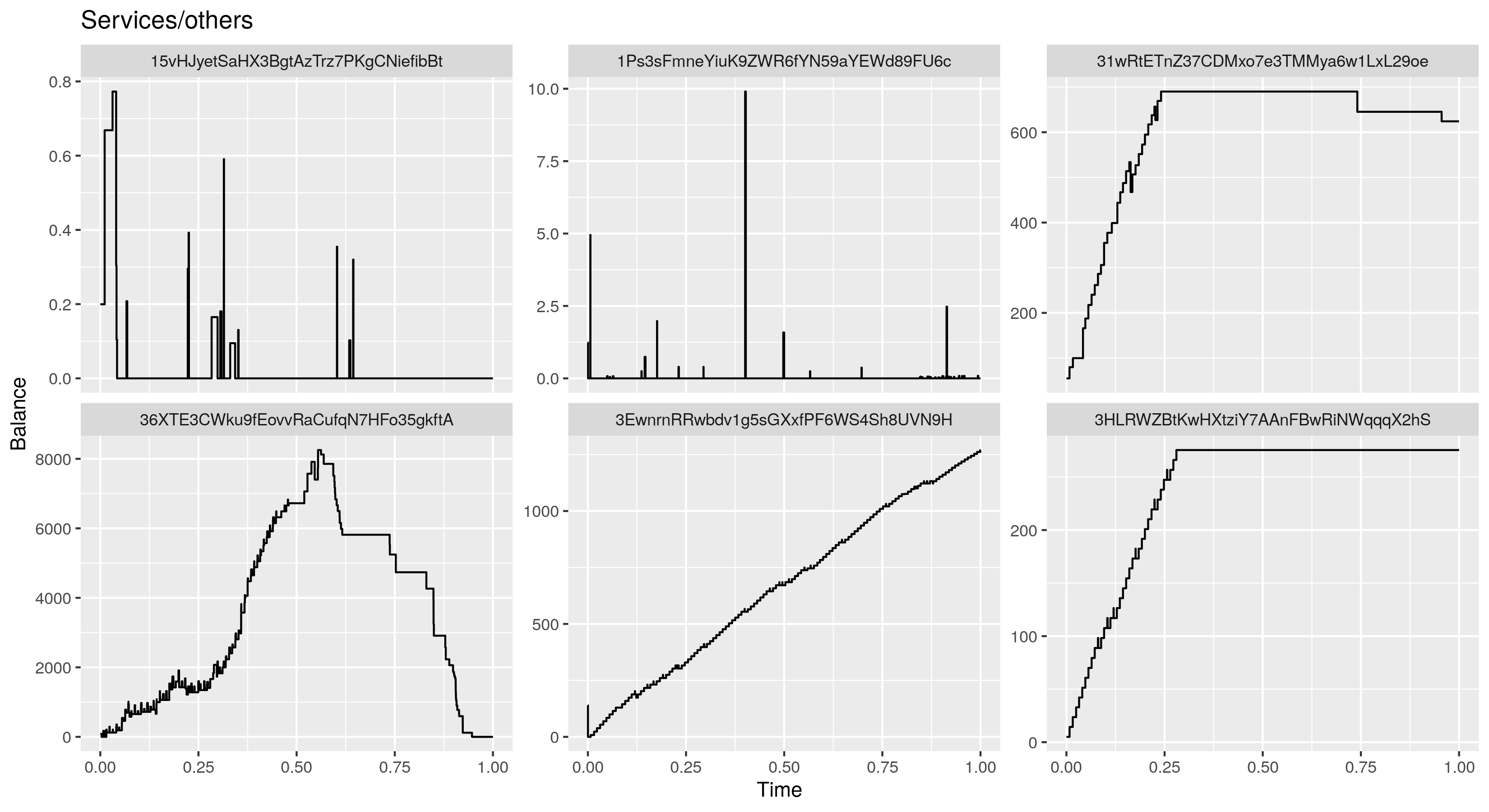}
    \caption{Original account balances for \final{six} services/others addresses, in dollars.}
    \label{fig:original_curves_services_others_20obs}
\end{figure}

\begin{figure}[H]
    \centering
    \includegraphics[scale = 0.55, center]{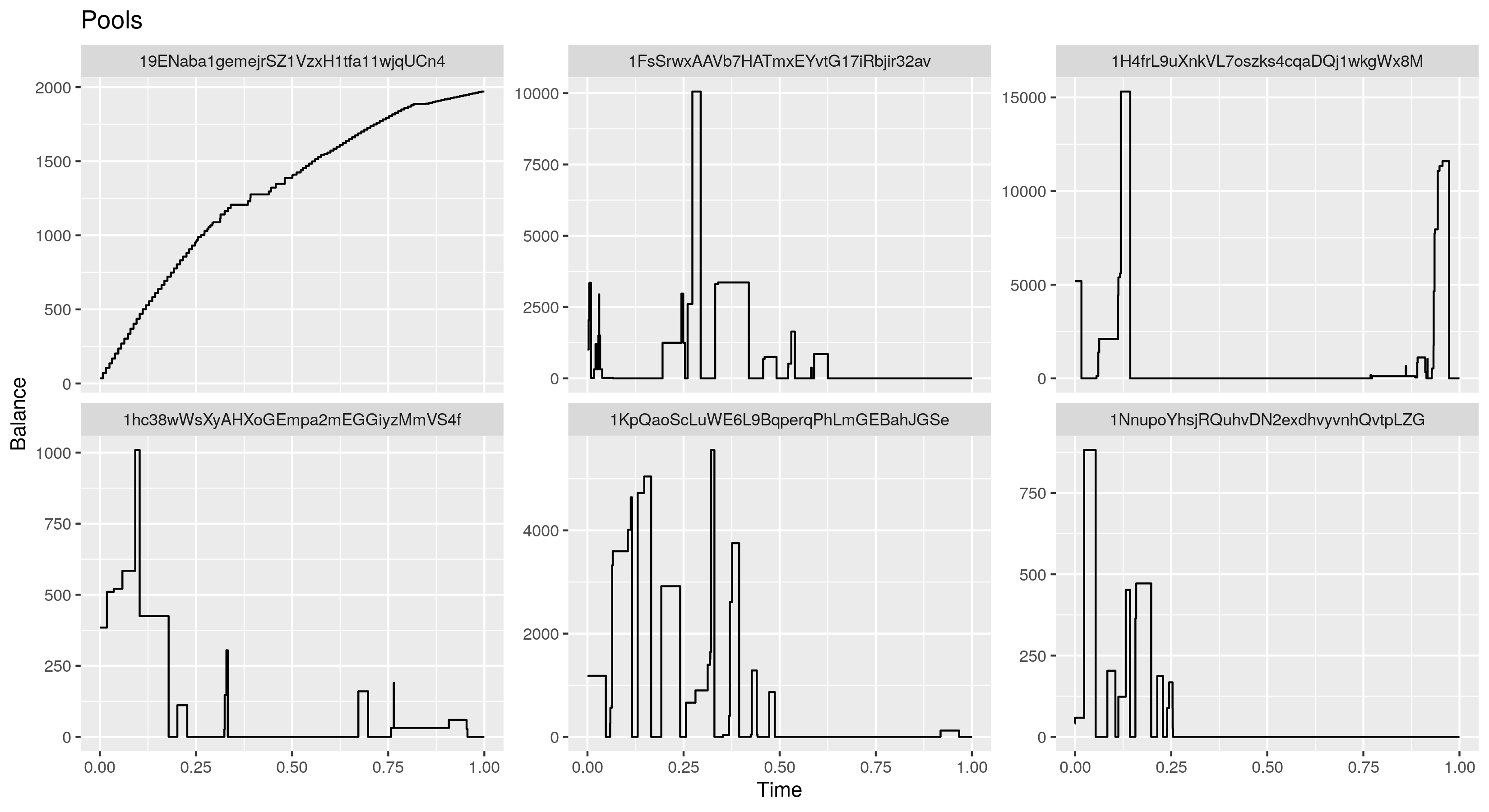}
    \caption{Original account balances for \final{six} pools addresses, in dollars.}
    \label{fig:original_curves_pools_20obs}
\end{figure}

\begin{figure}[H]
    \centering
    \includegraphics[scale = 0.55, center]{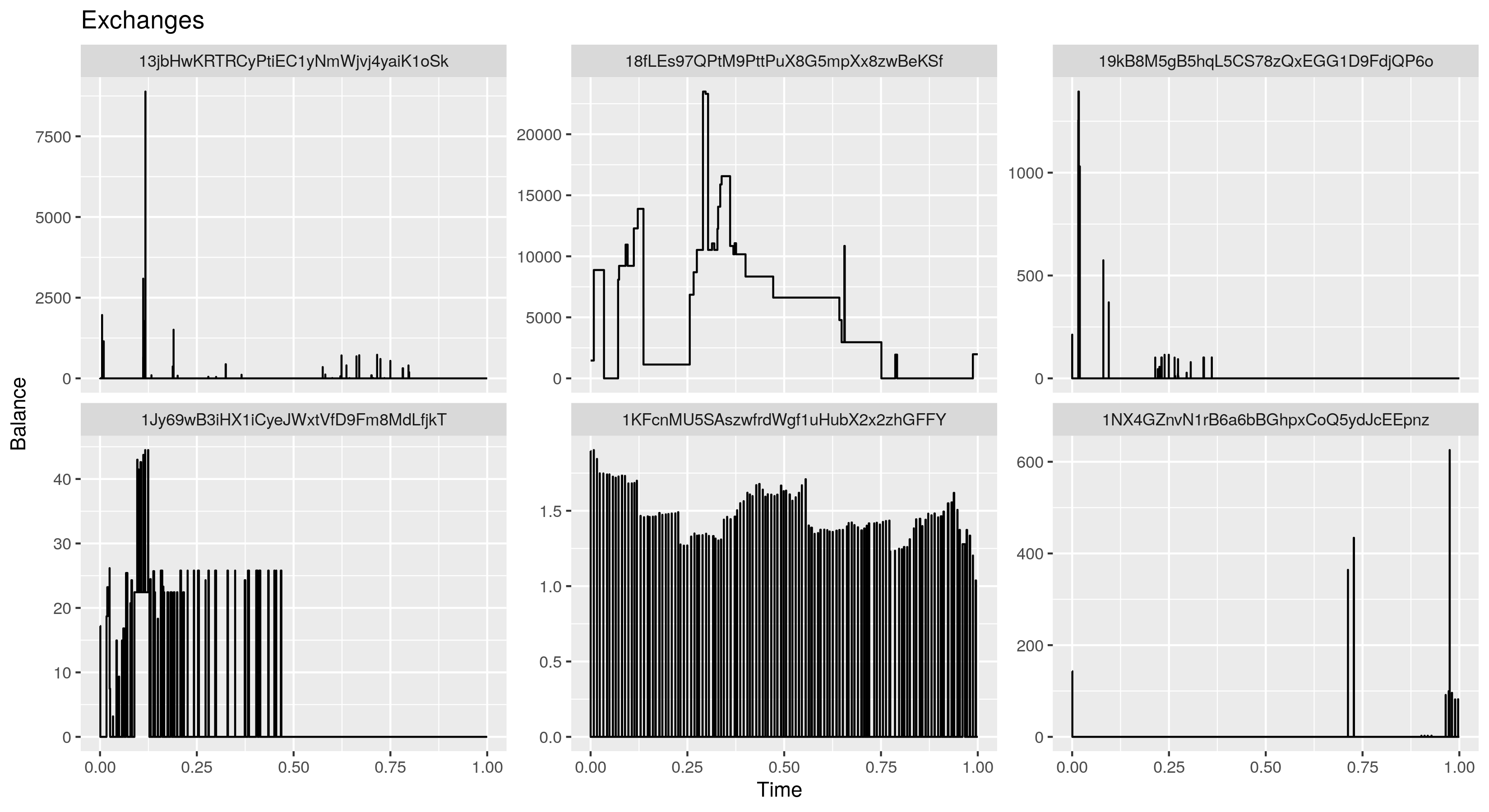}
    \caption{Original account balances for \final{six} exchange addresses, in dollars.}
    \label{fig:original_curves_exchange_20obs}
\end{figure}

\subsubsection{Accumulated credits and debits}

\begin{figure}[H]
    \centering
    \includegraphics[scale = 0.55, center]{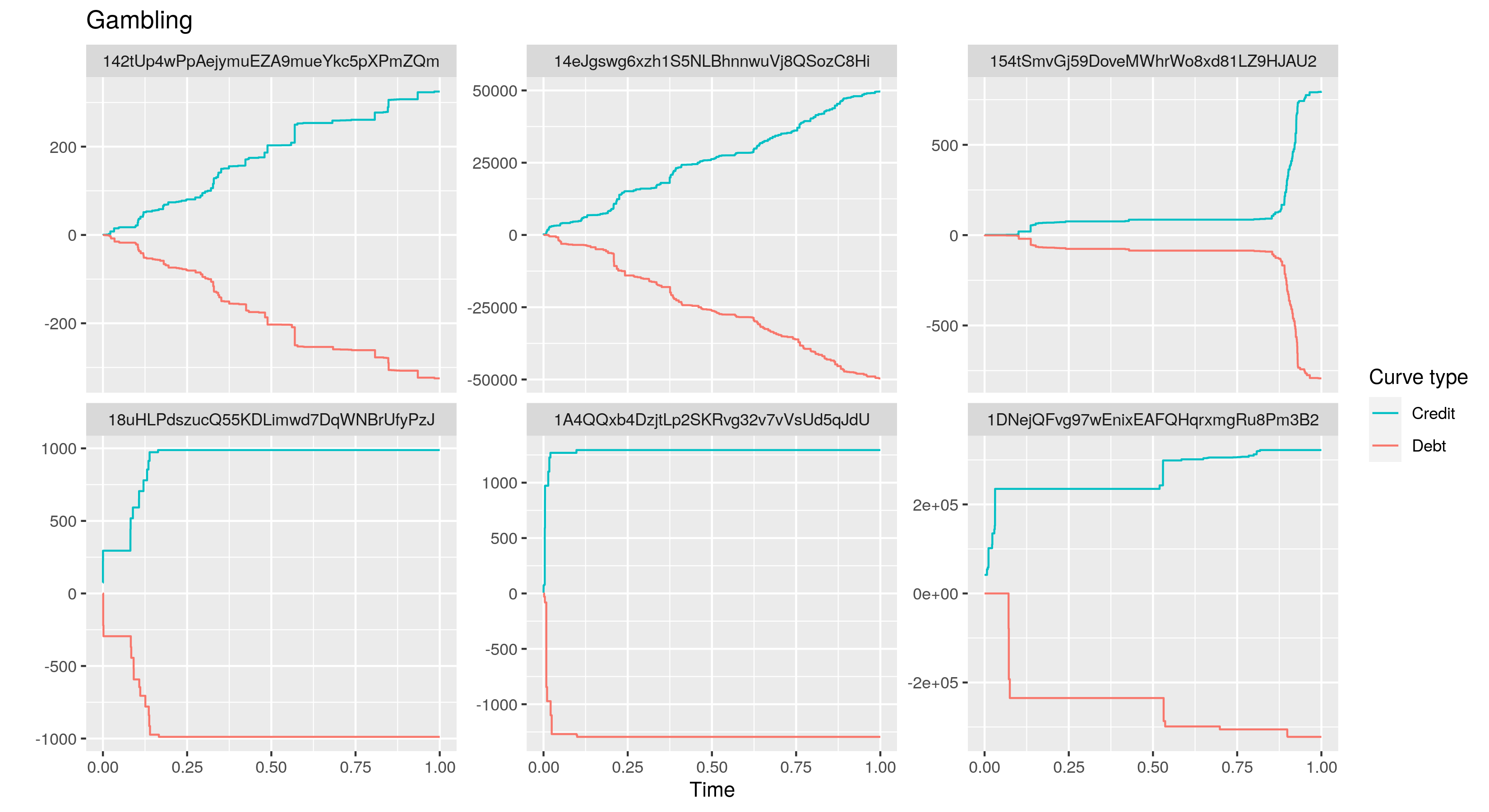}
    \caption{Accumulated sum of credits and debits for \final{six} gambling addresses, in dollars.}
    \label{fig:step_curves_gambling_20obs}
\end{figure}

\begin{figure}[H]
    \centering
    \includegraphics[scale = 0.55, center]{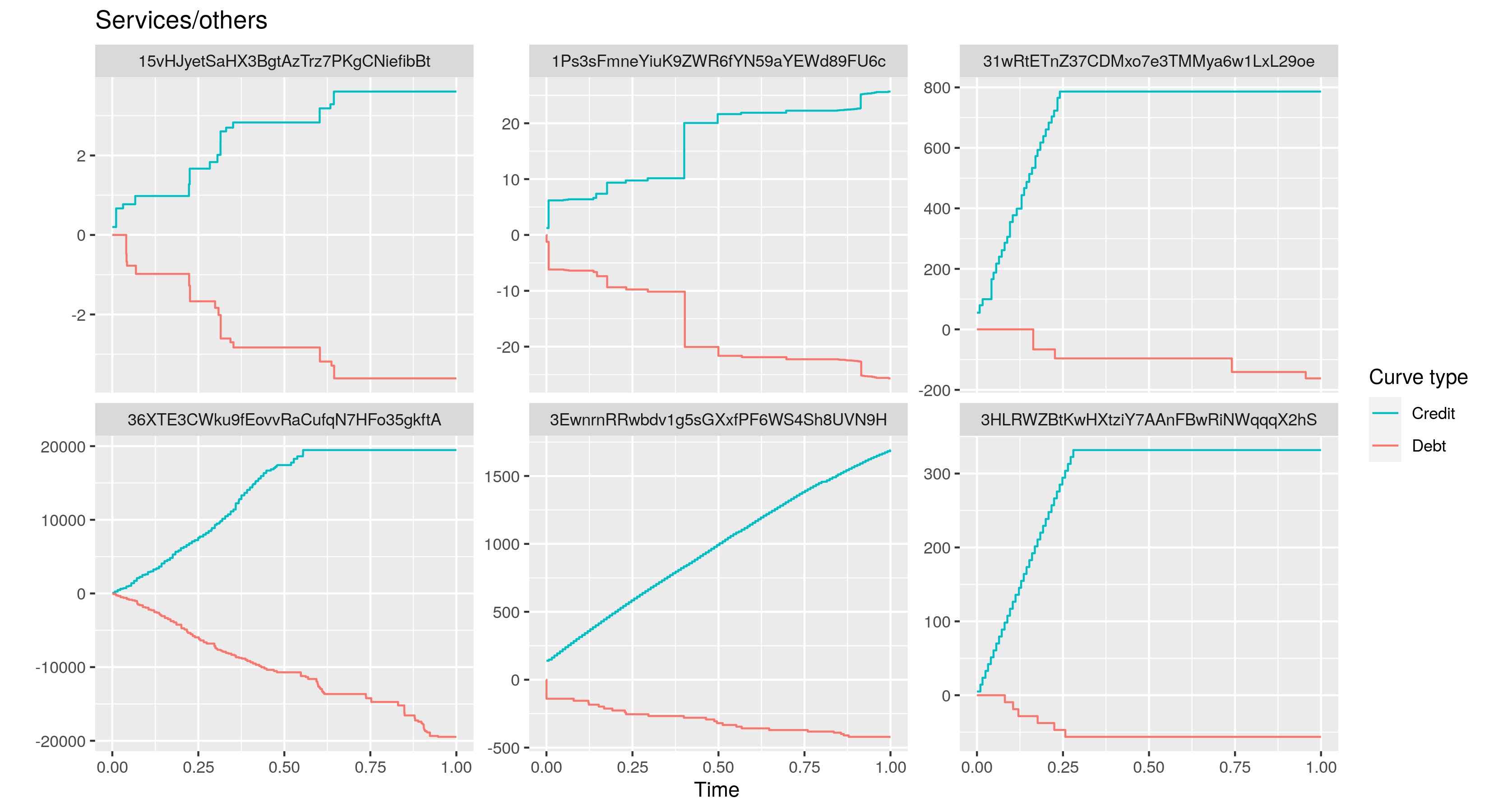}
    \caption{Accumulated sum of credits and debits for \final{six} services/others addresses, in dollars.}
    \label{fig:step_curves_services_other_20obs}
\end{figure}

\begin{figure}[H]
    \centering
    \includegraphics[scale = 0.55, center]{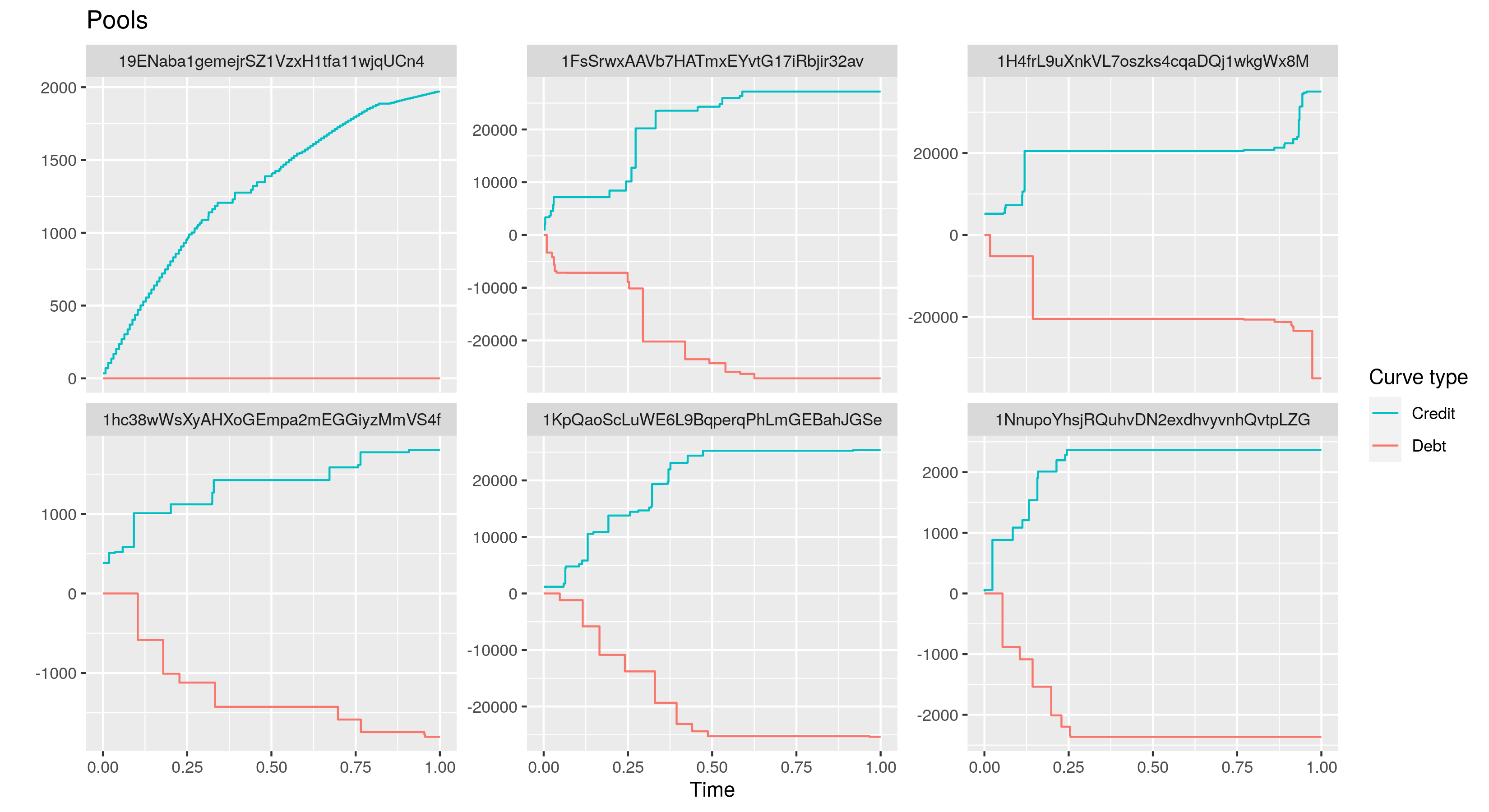}
    \caption{Accumulated sum of credits and debits for \final{six} pools addresses, in dollars.}
    \label{fig:step_curves_pools_20obs}
\end{figure}

\begin{figure}[H]
    \centering
    \includegraphics[scale = 0.55, center]{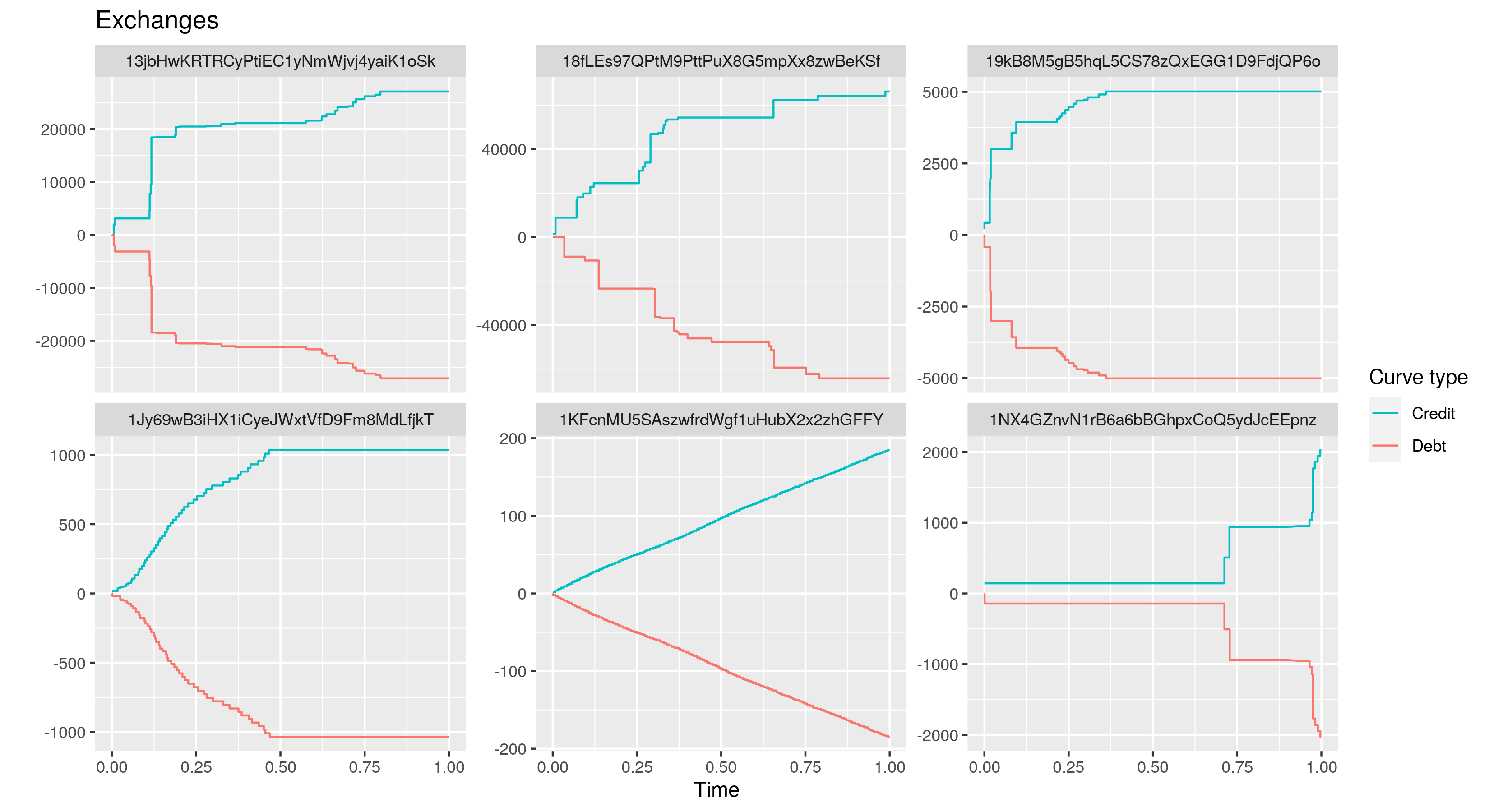}
    \caption{Accumulated sum of credits and debits for \final{six} exchange addresses, in dollars.}
    \label{fig:step_curves_exchange_20obs}
\end{figure}

\subsubsection{Smooth log-accumulated credits}

\begin{figure}[H]
    \centering
    \includegraphics[scale = 0.55, center]{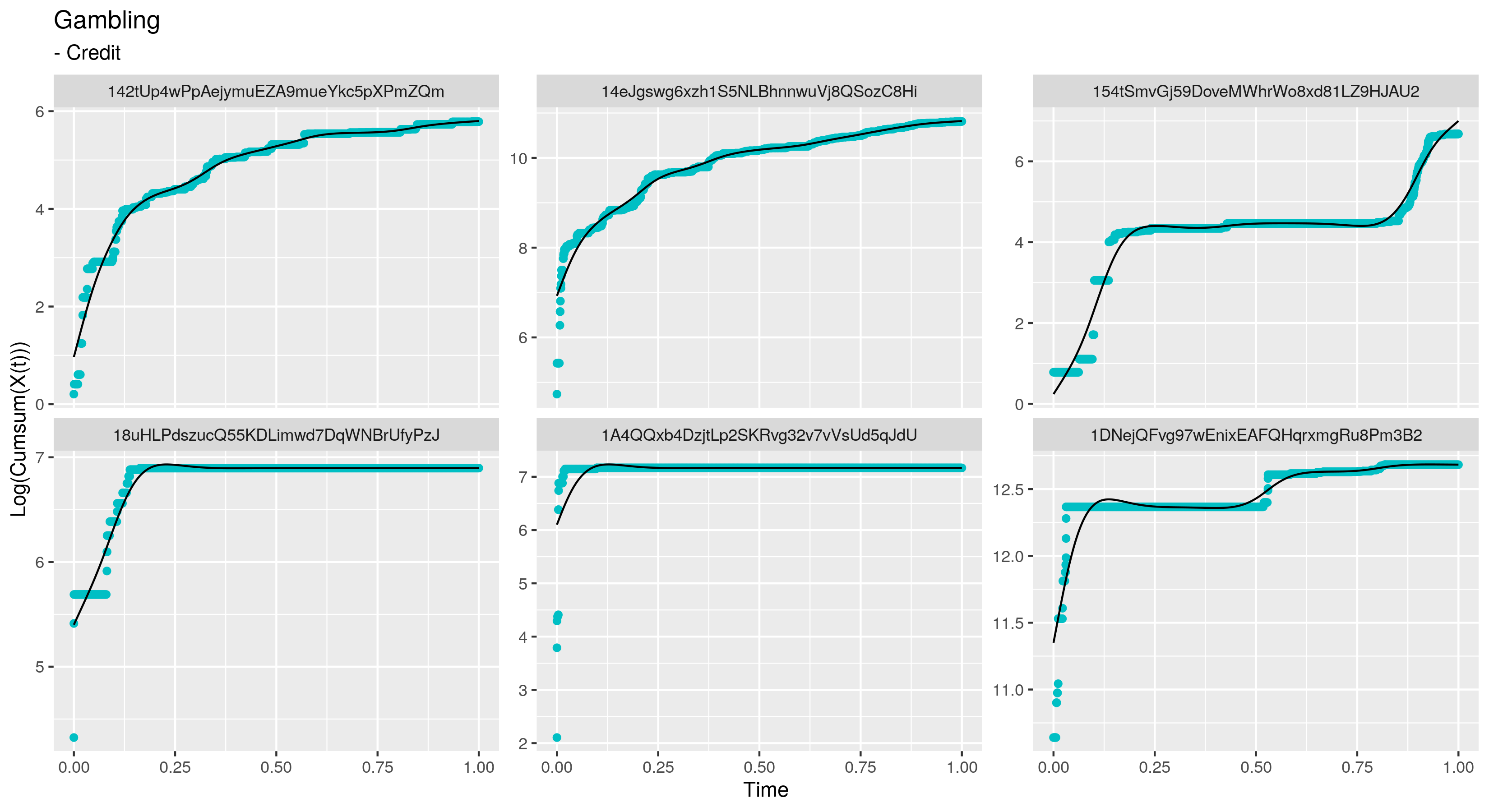}
    \caption{Smoothed log-accumulated sum of credits for \final{six} gambling addresses.}
    \label{fig:smooth_curves_gambling_20obs_credit}
\end{figure}

\begin{figure}[H]
    \centering
    \includegraphics[scale = 0.55, center]{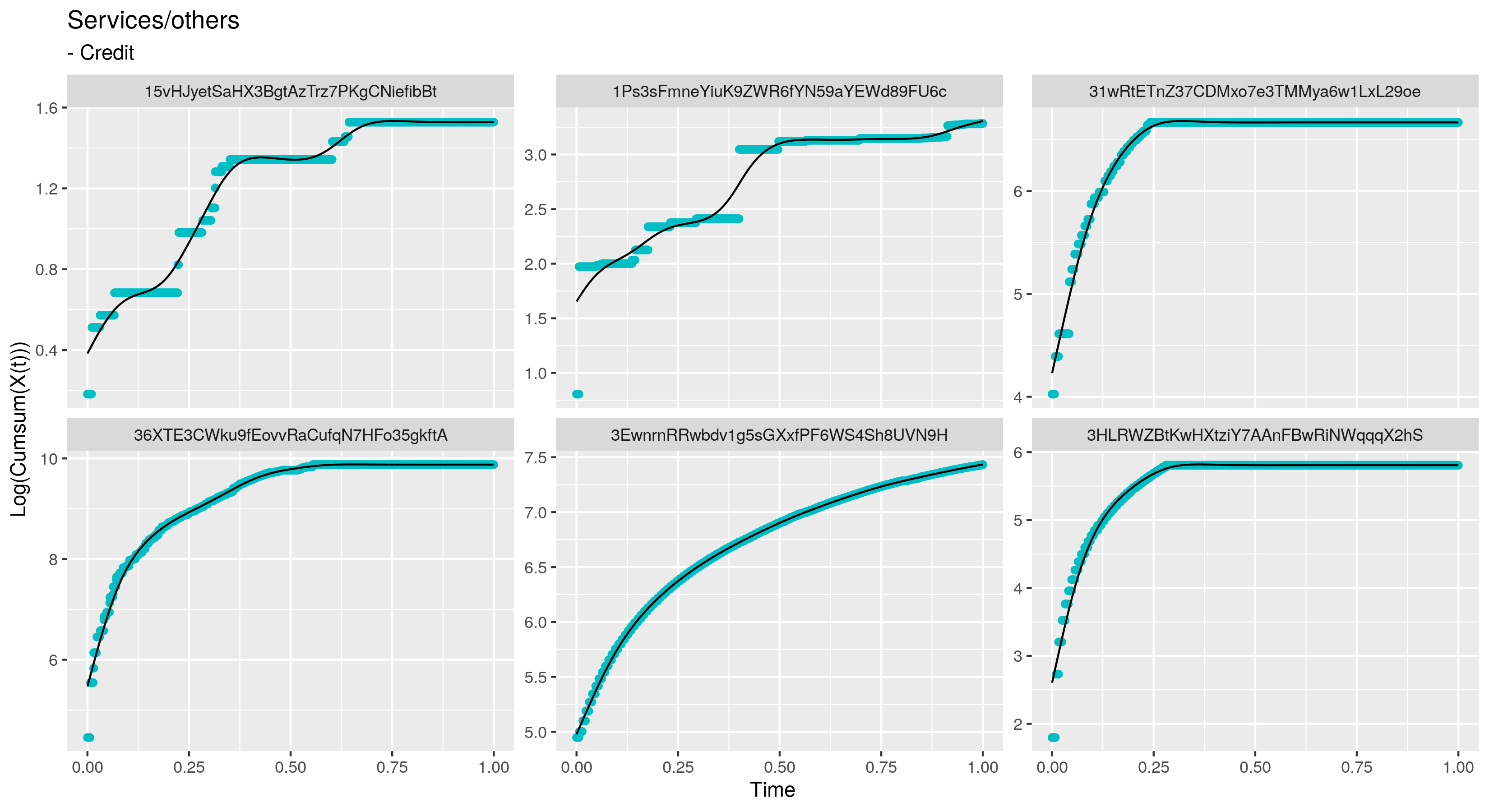}
    \caption{Smoothed log-accumulated sum of credits for \final{six} services/others addresses.}
    \label{fig:smooth_curves_services_others_20obs_credit}
\end{figure}

\begin{figure}[H]
    \centering
    \includegraphics[scale = 0.55, center]{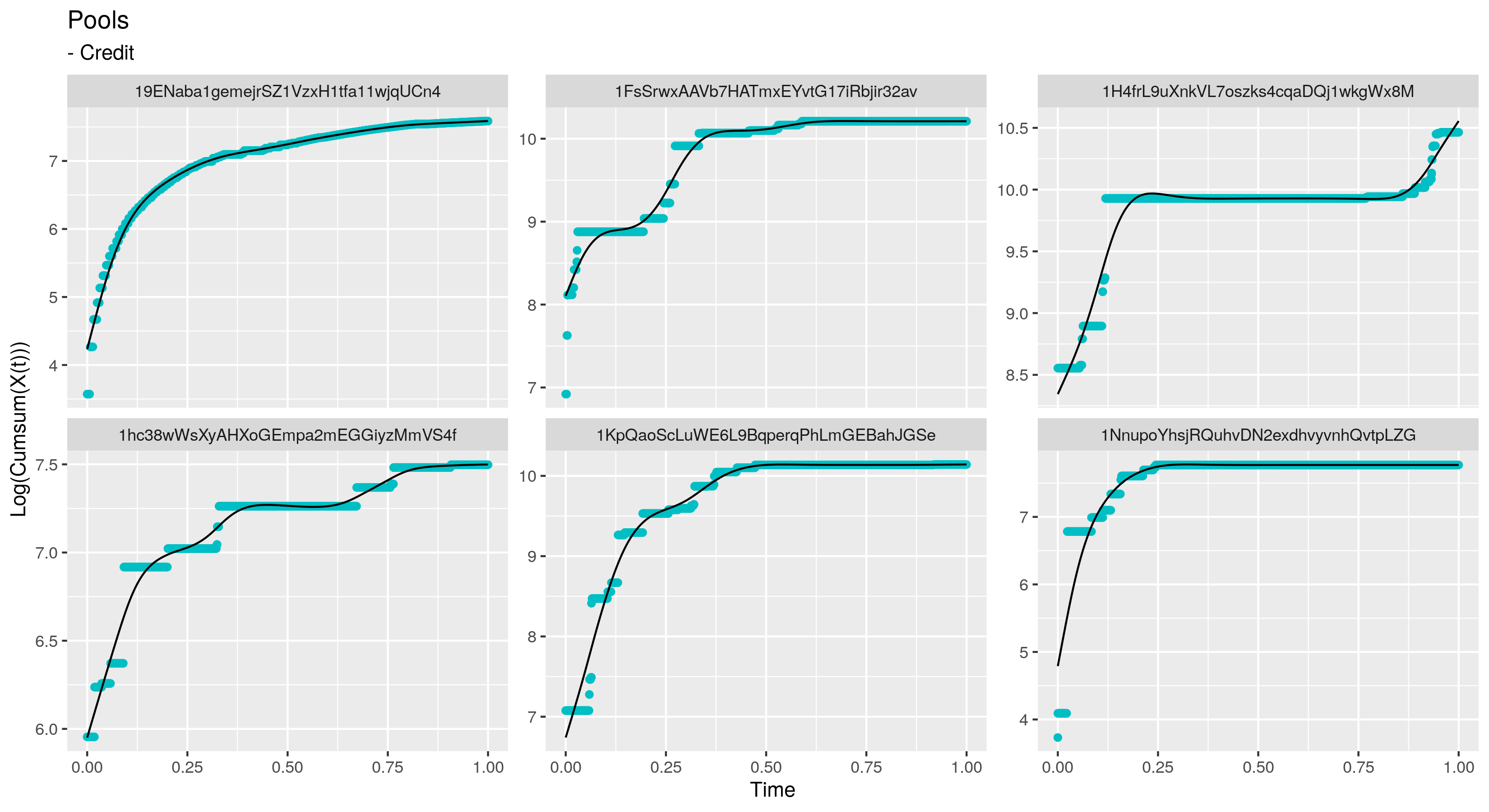}
    \caption{Smoothed log-accumulated sum of credits for \final{six} pools addresses.}
    \label{fig:smooth_curves_pools_20obs_credit}
\end{figure}

\begin{figure}[H]
    \centering
    \includegraphics[scale = 0.55, center]{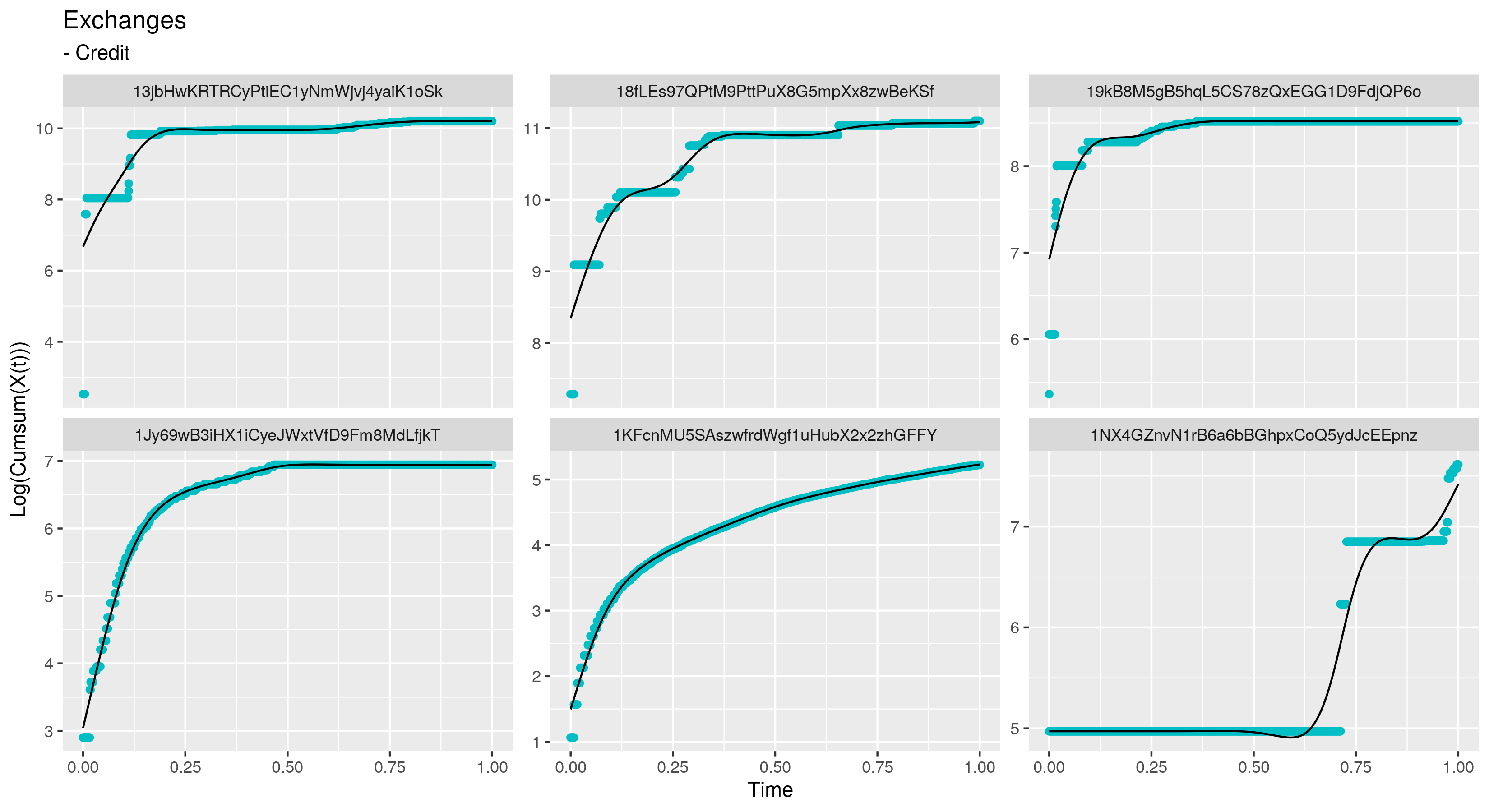}
    \caption{Smoothed log-accumulated sum of credits for \final{six} exchange addresses.}
    \label{fig:smooth_curves_exchange_20obs_credit}
\end{figure}

\subsubsection{Smooth derivatives of log-accumulated credits}

\begin{figure}[H]
    \centering
    \includegraphics[scale = 0.35, center]{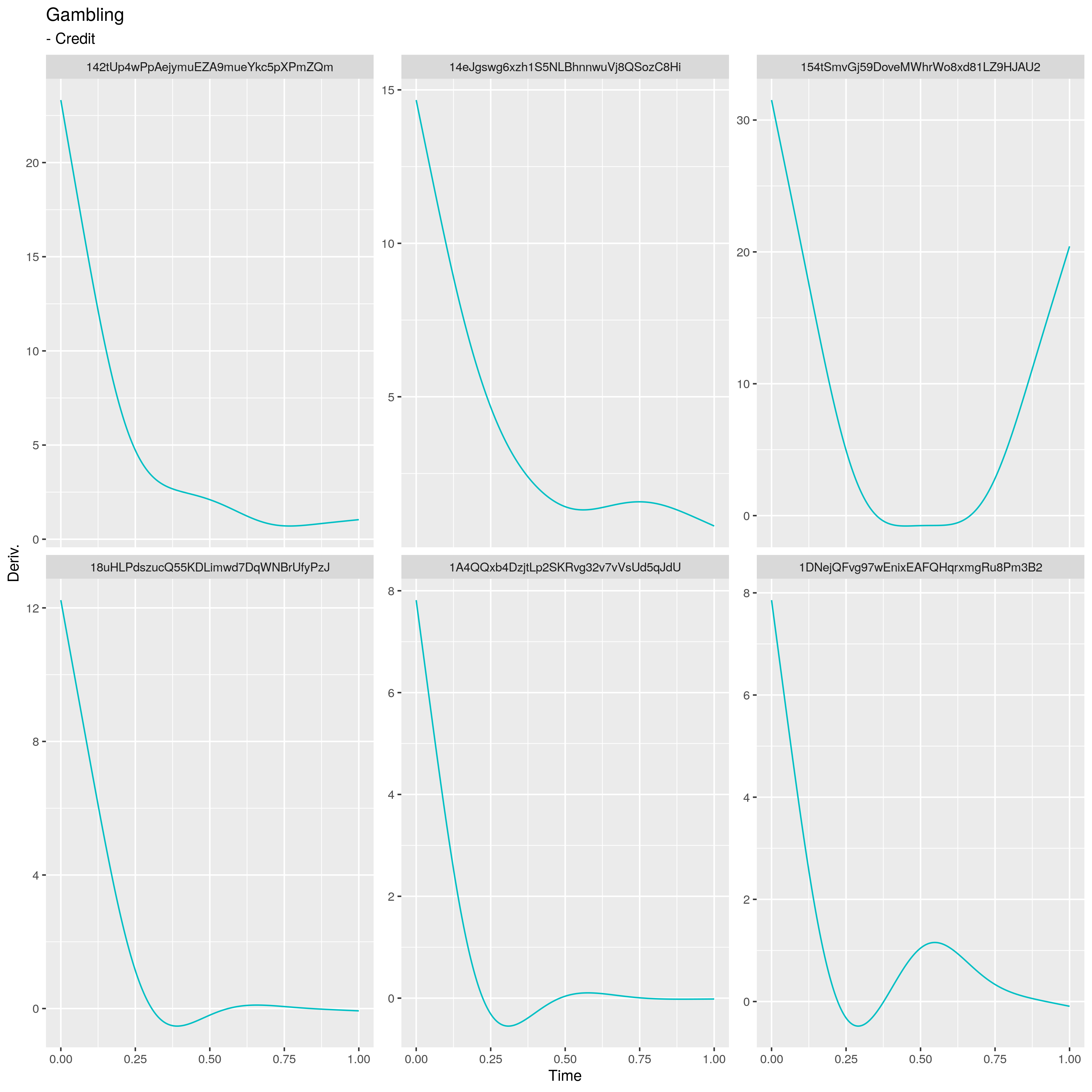}
    \caption{Derivatives of the smoothed log-accumulated sum of credits for \final{six} gambling addresses.}
    \label{fig:smooth_derivatives_gambling_20obs_credit}
\end{figure}

\begin{figure}[H]
    \centering
    \includegraphics[scale = 0.35, center]{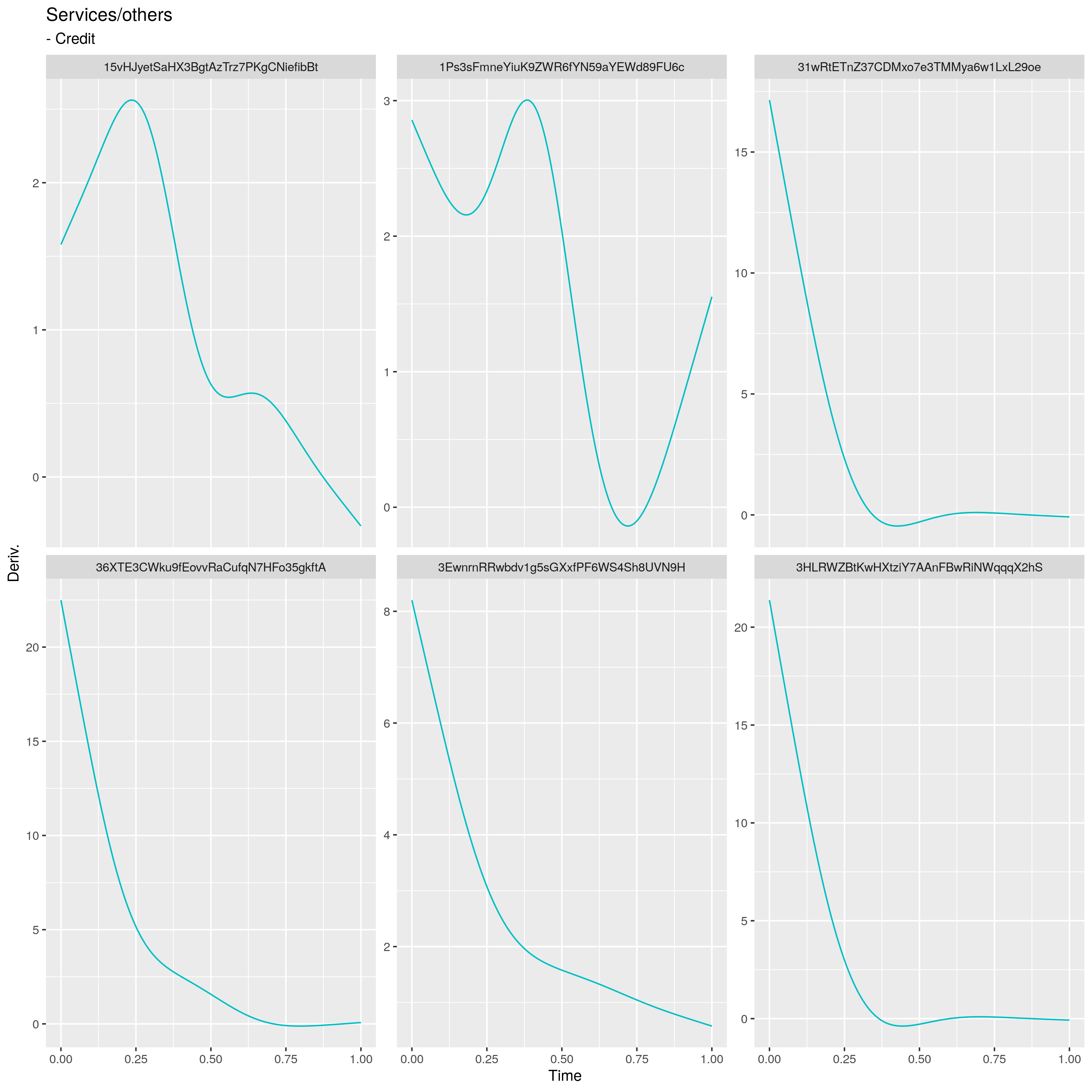}
    \caption{Derivatives of the smoothed log-accumulated sum of credits for \final{six} services/others addresses.}
    \label{fig:smooth_derivatives_services_others_20obs_credit}
\end{figure}

\begin{figure}[H]
    \centering
    \includegraphics[scale = 0.35, center]{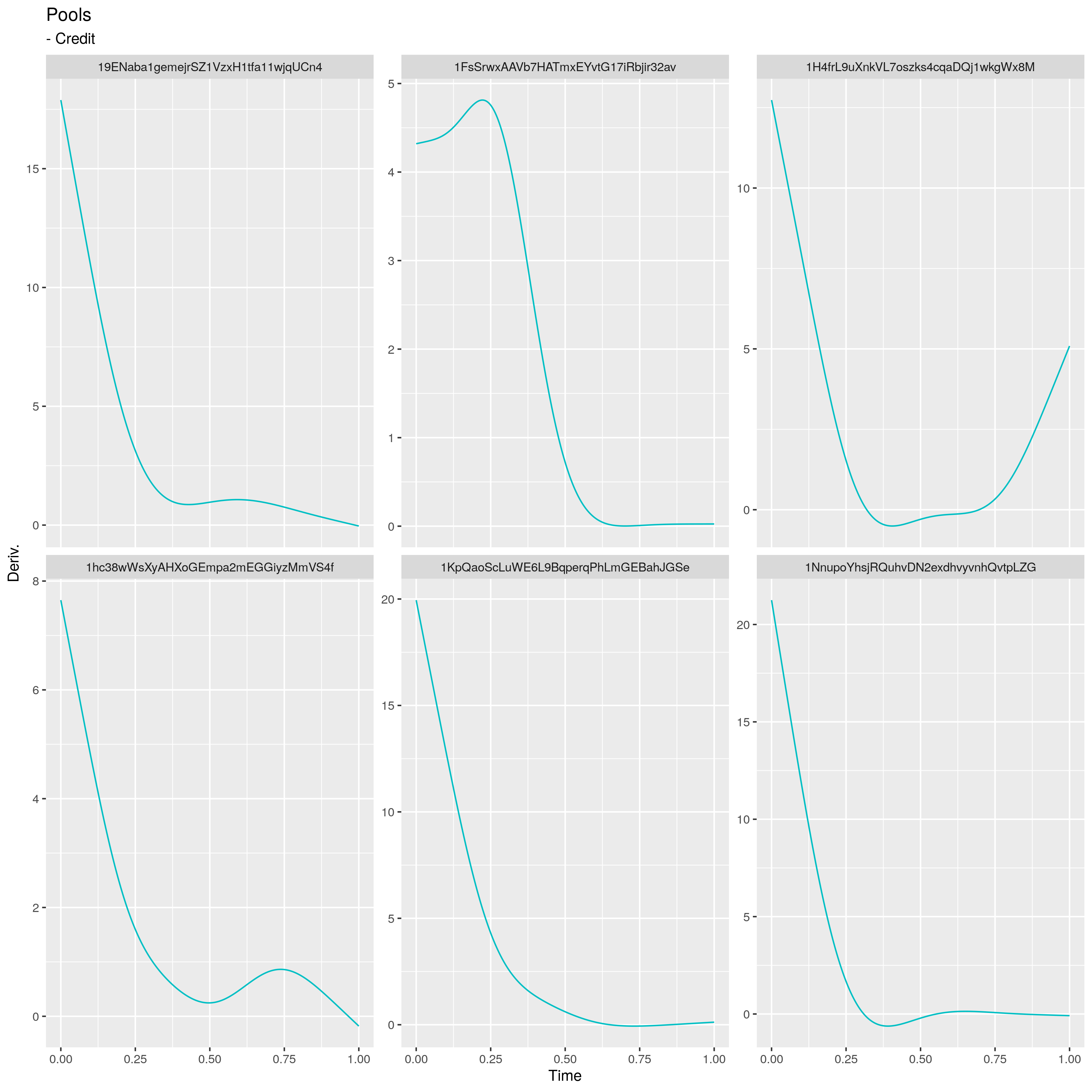}
    \caption{Derivatives of the smoothed log-accumulated sum of credits for \final{six} pools addresses.}
    \label{fig:smooth_derivatives_pools_20obs_credit}
\end{figure}

\begin{figure}[H]
    \centering
    \includegraphics[scale = 0.35, center]{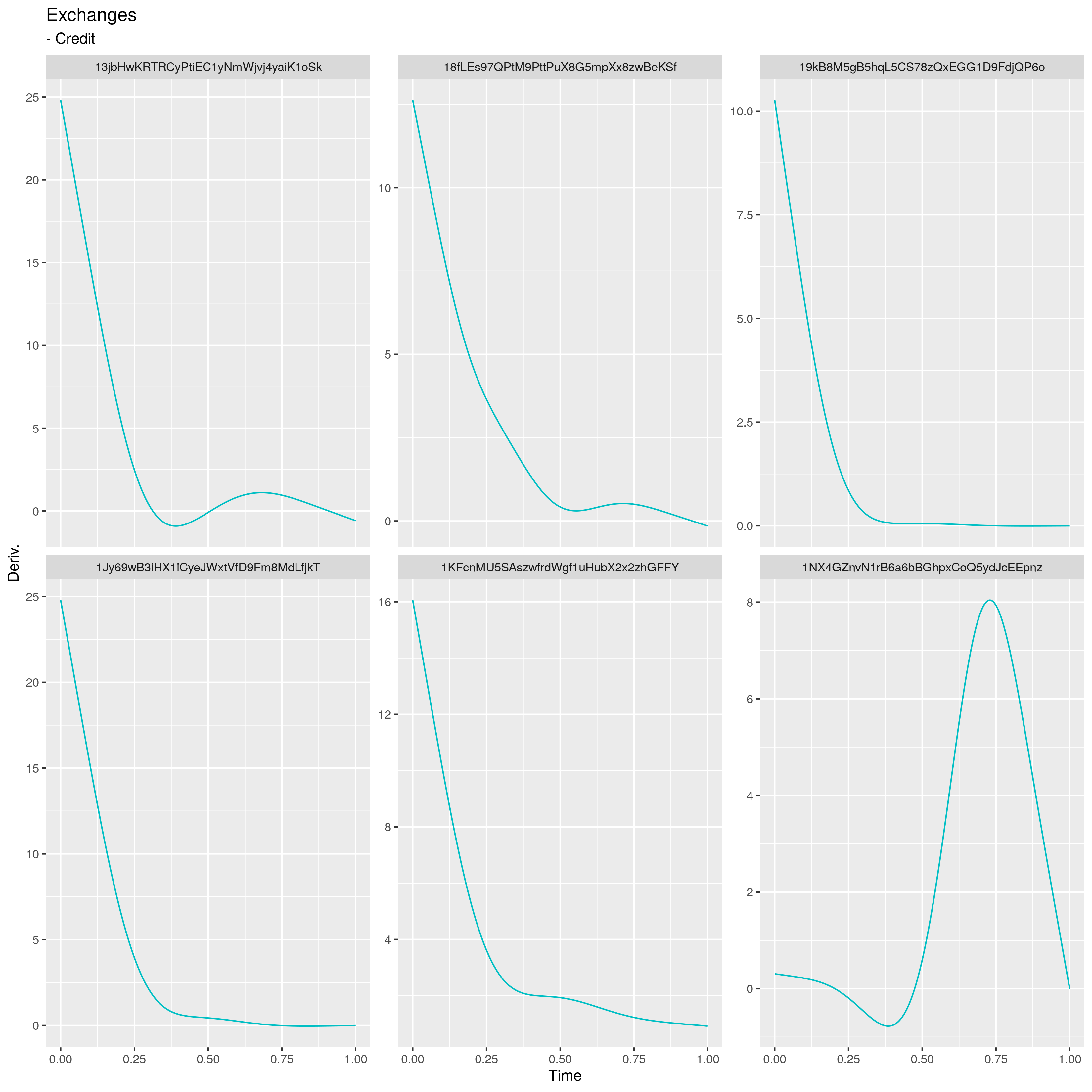}
    \caption{Derivatives of the smoothed log-accumulated sum of credits for \final{six} exchange addresses.}
    \label{fig:smooth_derivatives_exchange_20obs_credit}
\end{figure}

\subsubsection{Poisson rates}

\begin{figure}[H]
    \centering
    \includegraphics[scale = 0.55, center]{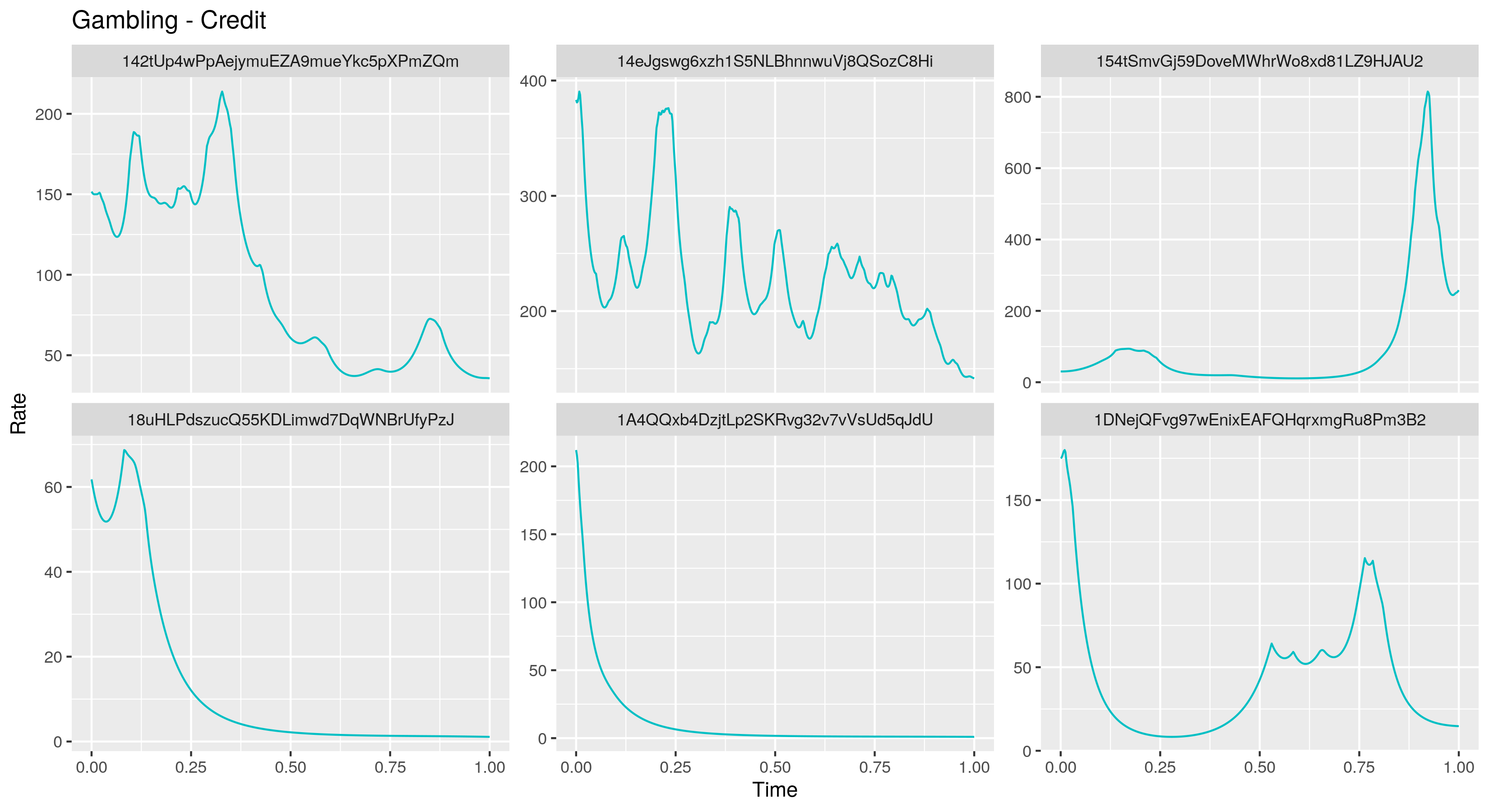}
    \caption{Poisson rate of credits for \final{six} gambling addresses.}
    \label{fig:poisson_rates_gambling_20obs_credit}
\end{figure}

\begin{figure}[H]
    \centering
    \includegraphics[scale = 0.55, center]{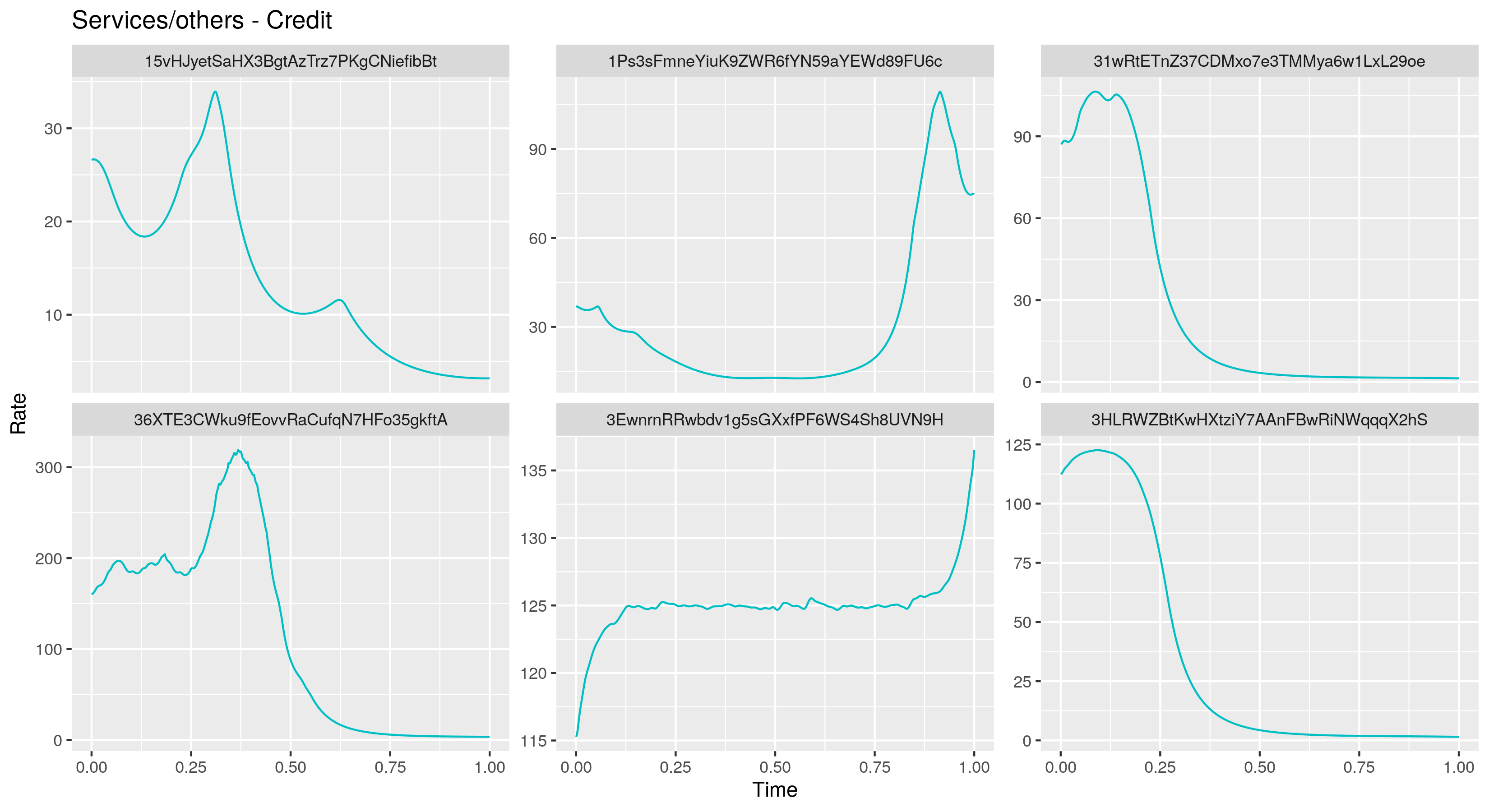}
    \caption{Poisson rate of credits for \final{six} services/others addresses.}
    \label{fig:poisson_rates_services_others_20obs_credit}
\end{figure}

\begin{figure}[H]
    \centering
    \includegraphics[scale = 0.55, center]{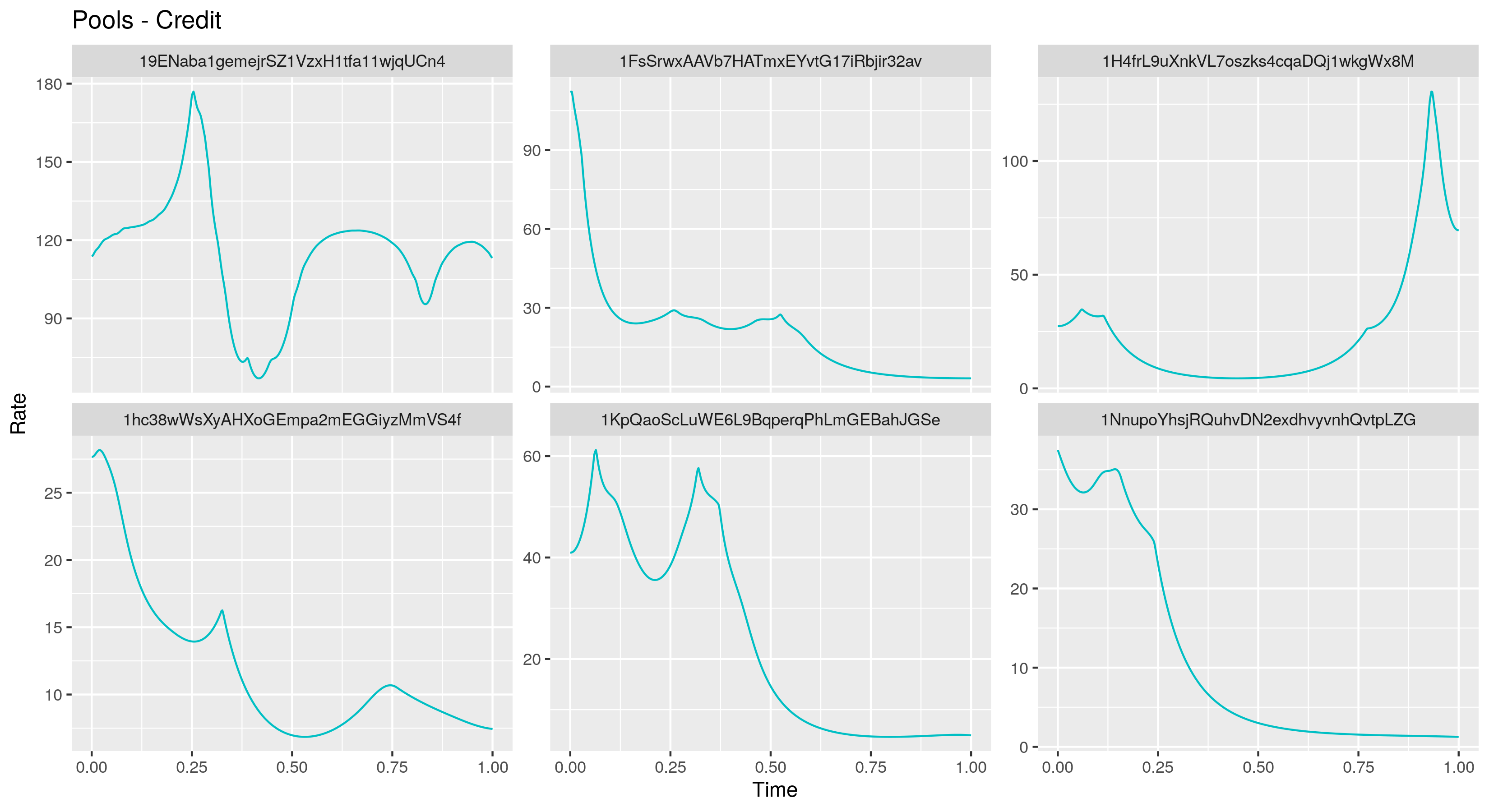}
    \caption{Poisson rate of credits for \final{six} pools addresses.}
    \label{fig:poisson_rates_pools_20obs_credit}
\end{figure}

\begin{figure}[H]
    \centering
    \includegraphics[scale = 0.55, center]{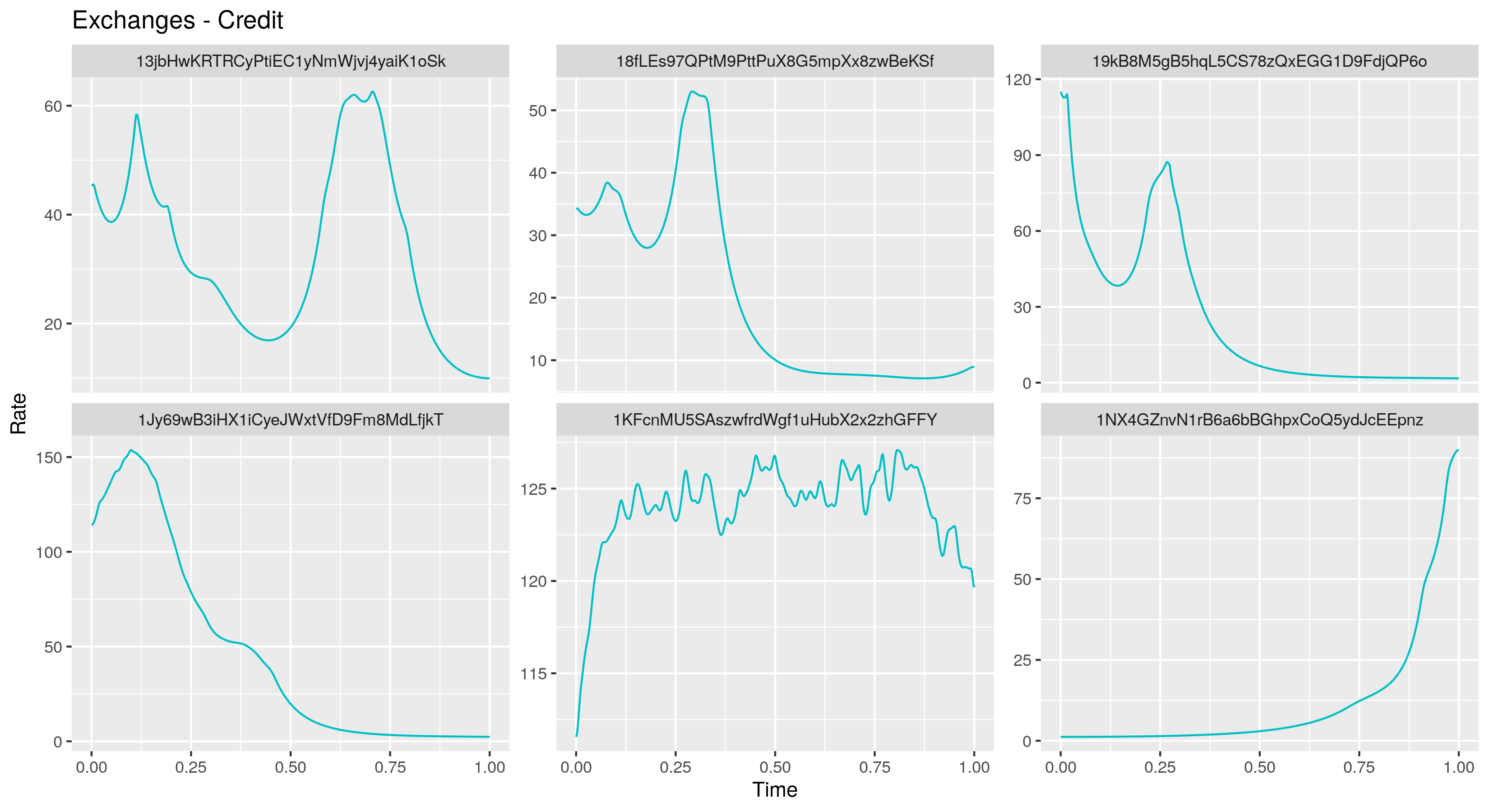}
    \caption{Poisson rate of credits for \final{six} exchange addresses.}
    \label{fig:poisson_rates_exchnge_20obs_credit}
\end{figure}

 \section{Functional Principal Components Analysis}
\label{sec:fda}

%Every observed data in the world is discrete. What makes discrete data functional, simply put, is the assumption that there is a function $x$ giving rise to the observed data. This underlying function should be, usually, smooth (otherwise, there is not much to gain by treating the data as functional instead of multivariate). Examples of functional data include height varying on age, temperature varying throughout the year, or financial data such as traded volume throughout the day. 

In this work, the account movements are assumed to be a function of an address' lifetime, therefore requiring the FDA methodology. For the sake of completeness, this appendix provides a brief theoretical background to the topic, focusing on the method employed in the article: functional principal components analysis (FPCA). The next sections are largely based on chapters 3 to 8 of \cite{RamsaySilverman2005book}.

Let, for $j \in \{ 1,\ldots,n\}$, where $n$ is the number of observations,
\begin{itemize}
    \item $y_j$ be the observation;
    \item $t_j$ be the time where $y_j$ is observed;
    \item $x_j = x(t_j)$ be the underlying, unknown function evaluated at time $t_j$;
    \item $\tau$ be a close interval of $\mathbb{R}$ where argument $t_j$ takes value, which in our case is $\tau = [0,1]$.
\end{itemize}

% We then assume 
% $$y_j = x_j + \epsilon_j,$$
% with $\epsilon_j \stackrel{\text{iid}}{\sim} \ \mathcal{N}(0, \sigma^2).$

We are interested in solving the following minimization problem: 
\begin{equation}
    \argmin_{x} \sum_{j=1}^n (y_j - x_j)^2 + \lambda \times PEN_m(x),
\end{equation}
where the roughness penalty is defined by the norm of the function's $m$-th derivative:
\begin{equation}
    \label{pen_m}
    PEN_m(x) = \int_{\tau} (D^m x(s))^2 ds.
\end{equation}

Let now $(\phi_k)_{k \in \mathbb{N}}$ be the pre-defined basis of $L^2[\tau]$. We can then approximate $x$ with a finite basis expansion, $x(t) = \sum_{k=1}^{\infty} c_k \phi_k(t) \approx \sum_{k=1}^K c_k \phi_k(t)$. In vector and matrix notation, we write $x(t) = \vec{c}^T \bphi(t)$, with $\bphi = [\phi_1,\ldots,\phi_K]^T$ and $\vec{c} = [c_1,\ldots,c_K]^T$, and the (approximated) minimization problem becomes:
\begin{equation}
    \label{argmin_basis}
    \hat{\vec{c}} = \argmin_{\vec{c}} \Big[ (\vec{y} - \vec{\Phi} \vec{c})^{T} (\vec{y} - \vec{\Phi} \vec{c}) \ + \lambda \times \vec{c}^T \vec{R} \vec{c} \Big] = (\vec{\Phi}^T \vec{\Phi} + \lambda\vec{R})^{-1} \vec{\Phi}^T \vec{y},
\end{equation}
where $\vec{R}_{K \times K} = \int_\tau D^m (\bphi(s))D^m (\bphi^T(s))ds$, $\vec{y}_{n \times 1}$ is the vector of observations and $\vec{\Phi}_{n \times K}$ is the matrix with entries $\phi_k(t_j)$. 

\subsection{Theoretical Results}

% CATAR REFERENCIAS DECENTES PROS TEOREMAS

The following theorem is a functional generalization of the Spectral Theorem for functions. First, some definitions:

\begin{definition}
We say $K: \tau \times \tau \to \mathbb{R}$ is a continuous, symmetric, non-negative kernel if $K$ is a continuous function, satisfying:
    \begin{equation}
            K(t, s) = K(s, t) \quad \text{and} \quad
            \sum_{i=1}^n \sum_{i=1}^n K(t_i, t_j) a_i a_j \geq 0,
    \end{equation}
    for all $t,s,t_1,\ldots t_n, \in \tau$ and $a_1,\ldots, a_n \in \mathbb{R}$, with $n \in \mathbb{N}$.
\end{definition}

\begin{definition}
The Hilbert-Schmidt operator associated to $K$ is defined on $L^2[\tau]$ as
    \begin{equation}
    \label{mercer1}
        (\mathcal{K} f) (t) = \int_{\tau} K(s,t)f(s)ds.
    \end{equation}
\end{definition}

\begin{theorem}[Mercer's Theorem]
Let $K$ be a continuous, symmetric, non-negative definite kernel. Then, there is a orthonormal basis $(\xi_i)_{i \in \mathbb{N}}$ of $L^2[\tau]$ consisting of the eigenfunctions of $\mathcal{K}$, such that the eigenvalues $(\rho_i)_{i \in \mathbb{N}}$ are non-negative. The eigenfunctions are continuous and $K$ can be written as: 
\begin{equation}
\label{mercer0}
    K(s,t) = \sum_{j=1}^\infty \rho_j \xi_j (s)\xi_j(t),
\end{equation}
where the convergence is absolute and uniform in $L^2[\tau]$. 
\end{theorem}

From Mercer's Theorem follows the eigenequation of the covariance operator $\mathcal{K}$:
\begin{align}
\label{mercer2}
        (\mathcal{K} \xi_i)(t) & = \int_{\tau}  K(s,t) \xi_i(s) ds 
        = \rho_i\xi_i(t).
\end{align}

This next theorem ensures that we can represent functional data with a principal components basis, as long as the function agrees to some assumptions.

\begin{theorem}[Karhunen–Loève Theorem]\label{thm:karhunen_loeve}
For a square-integrable\footnote{$\mathbb{E}\left [\int_{\tau} |X(t)|^2 dt \right] < \infty $} stochastic process $(X(t))_{t \in \tau}$, let $\mu(t) = \mathbb{E}[X(t)]$ be the mean function and $K(s,t) = Cov (X(s), X(t))$ be the covariance function. Then $K$ is a continuous, symmetric, non-negative definite kernel, and
\begin{equation}
\label{kl1}
    X(t) = \mu(t) + \sum_{j=1}^{\infty} Z_j \xi_j (t),
\end{equation}
where the convergence is uniform and in $L^2$. The random variables $(Z_j)_{j \in \mathbb{N}}$ are the functional principal components (FPCs). Moreover,
\begin{equation}
\label{kl2}
    Z_j = \int_{\tau} ( X(s) - \mu(s) )\xi_j(s) \, ds,
\end{equation}
with
\begin{equation}
    \mathbb{E}[Z_j] = 0, \quad  \mathbb{E}[Z_j Z_i] = 0 \, (i \neq j) \quad and \quad Var[Z_j] = \rho_j.
\end{equation}
\end{theorem}

From the above theorems, it is possible to conclude that eigenfunctions $\xi_i$ are uncorrelated and maximize the variance of the centered process $X(t) - \mu(t)$ projected on them.

\subsection{Estimating the functional principal components}
\label{estimating_fpca}

Although one could, if the functions are measured over a fine grid, use the discretization approach to find the FPCs, it would require the computation of the sample covariance matrix. We can avoid that by using the primary tool given by FDA: smoothing the data with pre-defined basis functions. 

First of all, we must center the functions. Supposing there are $M$ different functions, each function $x_i$ has a truncated basis expansion, with $(\phi_k)_{k \in \mathbb{N}}$ being any pre-available basis - the B-splines in our case: $x_i(t) \approx \sum_{k=1}^K a_{ik} \phi_k(t)$. The mean function can then be estimated simply by \begin{equation}
\label{eq:mu}
    \hat{\mu}(t) = \frac{1}{M} \sum_{i=1}^M  \sum_{k=1}^K a_{ik} \phi_k(t).
\end{equation}.

Now, for the sake of keeping the notation simple, let us denote $x_i$ as actually the centered function $x_i - \hat{\mu}$. Each centered function $x_i$ itself has a truncated basis expansion: $x_i(t) \approx \sum_{k=1}^K c_{ik} \phi_k(t)$.

We can then evaluate the function on sampling points in order to estimate $x_i$. Using vector notation, we write $\hat{\vec{x}}(s) = \vec{C} \bphi(s), \quad \vec{C} = (c_{ik})$ is the $M \times K$ matrix of coefficients. Notice that this representation of the coefficients in matrix $\vec{C}$ assumes that the curves were sampled at the same points. This matrix can be estimated using Equation (\ref{coef}).

The sample covariance function is then represented by:
\begin{equation}\label{eq:sample_cov}
    \hat{K}(s,t) = \frac{1}{M-1}\hat{\vec{x}}(s)^T \hat{\vec{x}}(t) = \frac{1}{M-1}\bphi(s)^T \vec{C}^T \vec{C} \bphi(t).
\end{equation}

Let us define the matrix $\vec{W}$ as the $K \times K$ symmetric matrix containing the inner products of the first $K$ basis components $\bphi = [\phi_1,\ldots,\phi_K]^T$: $w_{ij} = \langle \phi_i, \phi_j \rangle$, where $\langle \cdot, \cdot \rangle$ is the usual inner product in $L^2[\tau]$. Any eigenfunction of the Hilbert-Schmidt operator associated $\hat{K}$ can be written as an expansion of the basis $\bphi$:
\begin{equation}
\label{eq:eigein_expansion}
      \hat{\xi}(t) = \sum_{k=1}^K b_k \phi_k(t)  = \vec{b}^T\bphi(t).
\end{equation}
This yields, by Equation (\ref{eq:sample_cov}),
$$\int_{\tau} \hat{K}(s,t) \hat{\xi}(t) \, dt  = \frac{1}{M-1} \bphi(s)^T \vec{C}^T \vec{C} \vec{W} \vec{b}$$.

Hence, the eigenequation (\ref{mercer2}) for the sample covariance above can be expressed for the coefficients $\vec{b}$ as: 
\begin{equation}
\label{eq_eigen}
    \frac{1}{M-1}  \vec{C}^T \vec{C} \vec{W} \vec{b}  = \rho \vec{b}.
\end{equation}

However, the orthonormality required from the eigenfunctions $\hat{\xi}$'s implies that $||\hat{\xi}_i||^2 = 1$ and $\langle \hat{\xi}_i, \hat{\xi}_j \rangle = 0$, for $i\neq j$ in $\{1,\ldots,K\}$. The translation of the first restriction in terms of the basis coefficients is
\begin{align}
\label{eq:cond_ort1}
        ||\vec{b}_i^T\bphi||^2 &= 1 \implies \int_{\tau} \vec{b}_i^T \bphi(t) \bphi(t)^T \vec{b}_i \, dt  = \vec{b}_i^T \vec{W} \vec{b}_i = 1
\end{align}
The second restriction, in terms of basis coefficients, becomes
\begin{equation}
    \label{eq:cond_ort2}
        \langle \vec{b}_i^T \bphi , \vec{b}_j^T \bphi \rangle  = 0 
        \implies \int_{\tau} \vec{b}_i^T \bphi(t) \bphi(t)^T \vec{b}_j \, dt  = \vec{b}^T_i \vec{W} \vec{b}_j = 0.
\end{equation}
These restrictions shown in Equations (\ref{eq:cond_ort1}) and (\ref{eq:cond_ort2}) are not trivially satisfied, since solving the eigenproblem in Equation (\ref{eq_eigen}) gives orthonormality for $\vec{b}_i$'s instead. However, the restrictions can be accounted for by defining $\vec{u}_i = \vec{W}^{\frac{1}{2}} \vec{b}_i$ and multiplying Equation (\ref{eq_eigen}) by $\vec{W}^{\frac{1}{2}}$ on the left side in order to build the equivalent eigenproblem:

\begin{equation}
\label{eq_eigen_u}
     \frac{1}{M-1} \vec{W}^{\frac{1}{2}} \vec{C}^T \vec{C} \vec{W}^{\frac{1}{2}} \vec{u}_i  = \rho_i \vec{u}_i.
\end{equation}

Equation (\ref{eq_eigen_u}) can be easily solved for $\vec{u}_i$ using any linear algebra package, which returns eigenvectors $\vec{u}_i$'s such that $\vec{u}_i^T \vec{u}_i = 1$ and $\langle \vec{u}_i, \vec{u}_j \rangle = 0$, and because $\vec{W}$ is symmetric and by the definition of $\vec{u}_i$, it is easy to see that the conditions on $\vec{b}_i$ are satisfied.

%One interesting case that deserves mentioning is when $\vec{W}$ is the identity matrix (which occurs, for example, when the Fourier basis is chosen as the pre-defined basis). When this happens, the eigenanalysis reduces to performing standard multivariate PCA on the coefficient matrix $\vec{C}$, and normalizing by $N-1$.

At last, it should be highlighted that performing FPCA in its classical version, described above, requires curves sampled at the same points. Although there are a few techniques perform FPCA for irregularly sampled points \cite{yao2005functional}, they are not in the scope of this work.


\begin{thebibliography}{}

\bibitem[\protect\citeauthoryear{Aneiros, Novo, and Vieu}{Aneiros
  et~al.}{2021}]{aneiros2021variable}
Aneiros, G., S.~Novo, and P.~Vieu (2021).
\newblock Variable selection in functional regression models: A review.
\newblock {\em Journal of Multivariate Analysis\/}, 104871.

\bibitem[\protect\citeauthoryear{Cuevas, Febrero, and Fraiman}{Cuevas
  et~al.}{2007}]{cuevas2007robust}
Cuevas, A., M.~Febrero, and R.~Fraiman (2007).
\newblock Robust estimation and classification for functional data via
  projection-based depth notions.
\newblock {\em Computational Statistics\/}~{\em 22\/}(3), 481--496.

\bibitem[\protect\citeauthoryear{Escabias, Aguilera, and Valderrama}{Escabias
  et~al.}{2004}]{escabias2004principal}
Escabias, M., A.~Aguilera, and M.~Valderrama (2004).
\newblock Principal component estimation of functional logistic regression:
  discussion of two different approaches.
\newblock {\em Journal of Nonparametric Statistics\/}~{\em 16\/}(3-4),
  365--384.

\bibitem[\protect\citeauthoryear{Escabias, Aguilera, and Valderrama}{Escabias
  et~al.}{2005}]{escabias2005modeling}
Escabias, M., A.~Aguilera, and M.~Valderrama (2005).
\newblock Modeling environmental data by functional principal component
  logistic regression.
\newblock {\em Environmetrics: The official journal of the International
  Environmetrics Society\/}~{\em 16\/}(1), 95--107.

\bibitem[\protect\citeauthoryear{Febrero-Bande and
  Gonz{\'a}lez-Manteiga}{Febrero-Bande and
  Gonz{\'a}lez-Manteiga}{2013}]{febrero2013generalized}
Febrero-Bande, M. and W.~Gonz{\'a}lez-Manteiga (2013).
\newblock Generalized additive models for functional data.
\newblock {\em Test\/}~{\em 22\/}(2), 278--292.

\bibitem[\protect\citeauthoryear{Febrero-Bande, González-Manteiga, Prallon,
  and Saporito}{Febrero-Bande et~al.}{2022}]{fda_bitcoin_extended}
Febrero-Bande, M., W.~González-Manteiga, B.~Prallon, and Y.~Saporito (2022).
\newblock Functional classification of bitcoin addresses.
\newblock {\em Preprint\/}.
\newblock Available at arXiv: https://arxiv.org/abs/2202.12019.

\bibitem[\protect\citeauthoryear{Febrero-Bande and {Oviedo de la
  Fuente}}{Febrero-Bande and {Oviedo de la Fuente}}{2012}]{fdauscpackage}
Febrero-Bande, M. and M.~{Oviedo de la Fuente} (2012).
\newblock Statistical computing in functional data analysis: The {R} package
  {fda.usc}.
\newblock {\em Journal of Statistical Software\/}~{\em 51\/}(4), 1--28.

\bibitem[\protect\citeauthoryear{Foley, Karlsen, and Putni{\c{n}}{\v{s}}}{Foley
  et~al.}{2019}]{foley2019sex}
Foley, S., J.~R. Karlsen, and T.~J. Putni{\c{n}}{\v{s}} (2019).
\newblock Sex, drugs, and bitcoin: How much illegal activity is financed
  through cryptocurrencies?
\newblock {\em The Review of Financial Studies\/}~{\em 32\/}(5), 1798--1853.

\bibitem[\protect\citeauthoryear{Friedman, Hastie, and Tibshirani}{Friedman
  et~al.}{2001}]{friedman2001elements}
Friedman, J., T.~Hastie, and R.~Tibshirani (2001).
\newblock {\em The elements of statistical learning}.
\newblock Springer series in statistics New York.

\bibitem[\protect\citeauthoryear{Hall and Hosseini-Nasab}{Hall and
  Hosseini-Nasab}{2006}]{hall2006properties}
Hall, P. and M.~Hosseini-Nasab (2006).
\newblock On properties of functional principal components analysis.
\newblock {\em Journal of the Royal Statistical Society: Series B (Statistical
  Methodology)\/}~{\em 68\/}(1), 109--126.

\bibitem[\protect\citeauthoryear{Hall, Poskitt, and Presnell}{Hall
  et~al.}{2001}]{hall2001functional}
Hall, P., D.~S. Poskitt, and B.~Presnell (2001).
\newblock A functional data—analytic approach to signal discrimination.
\newblock {\em Technometrics\/}~{\em 43\/}(1), 1--9.

\bibitem[\protect\citeauthoryear{Hu, Seneviratne, Thilakarathna, Fukuda, and
  Seneviratne}{Hu et~al.}{2019}]{hu2019characterizing}
Hu, Y., S.~Seneviratne, K.~Thilakarathna, K.~Fukuda, and A.~Seneviratne (2019).
\newblock Characterizing and detecting money laundering activities on the
  bitcoin network.
\newblock {\em arXiv preprint arXiv:1912.12060\/}.

\bibitem[\protect\citeauthoryear{Jacobsen and Gani}{Jacobsen and
  Gani}{2006}]{jacobsen2006point}
Jacobsen, M. and J.~Gani (2006).
\newblock {\em Point process theory and applications: marked point and
  piecewise deterministic processes}.
\newblock Springer.

\bibitem[\protect\citeauthoryear{Jiang and Chen}{Jiang and
  Chen}{2020}]{jiang2020filtering}
Jiang, C.-R. and L.-H. Chen (2020).
\newblock Filtering-based approaches for functional data classification.
\newblock {\em Wiley Interdisciplinary Reviews: Computational
  Statistics\/}~{\em 12\/}(4), e1490.

\bibitem[\protect\citeauthoryear{Jourdan, Blandin, Wynter, and
  Deshpande}{Jourdan et~al.}{2018}]{jourdan2018characterizing}
Jourdan, M., S.~Blandin, L.~Wynter, and P.~Deshpande (2018).
\newblock Characterizing entities in the bitcoin blockchain.
\newblock In {\em 2018 IEEE International Conference on Data Mining Workshops
  (ICDMW)}, pp.\  55--62. IEEE.

\bibitem[\protect\citeauthoryear{Lee}{Lee}{2004}]{lee2004functional}
Lee, H.-J. (2004).
\newblock {\em Functional data analysis: classification and regression}.
\newblock Texas A\&M University.

\bibitem[\protect\citeauthoryear{Leng and M{\"u}ller}{Leng and
  M{\"u}ller}{2006}]{leng2006classification}
Leng, X. and H.-G. M{\"u}ller (2006).
\newblock Classification using functional data analysis for temporal gene
  expression data.
\newblock {\em Bioinformatics\/}~{\em 22\/}(1), 68--76.

\bibitem[\protect\citeauthoryear{Li, Xiao, Xia, Tang, and Li}{Li
  et~al.}{2013}]{li2013hyperspectral}
Li, H., G.~Xiao, T.~Xia, Y.~Y. Tang, and L.~Li (2013).
\newblock Hyperspectral image classification using functional data analysis.
\newblock {\em IEEE transactions on Cybernetics\/}~{\em 44\/}(9), 1544--1555.

\bibitem[\protect\citeauthoryear{Li, Qiu, and Xu}{Li
  et~al.}{2022}]{li2022multivariate}
Li, Y., Y.~Qiu, and Y.~Xu (2022).
\newblock From multivariate to functional data analysis: Fundamentals, recent
  developments, and emerging areas.
\newblock {\em Journal of Multivariate Analysis\/}~{\em 188}, 104806.

\bibitem[\protect\citeauthoryear{Ling and Vieu}{Ling and
  Vieu}{2021}]{ling2021semiparametric}
Ling, N. and P.~Vieu (2021).
\newblock On semiparametric regression in functional data analysis.
\newblock {\em Wiley Interdisciplinary Reviews: Computational
  Statistics\/}~{\em 13\/}(6), e1538.

\bibitem[\protect\citeauthoryear{L{\'o}pez-Pintado and Romo}{L{\'o}pez-Pintado
  and Romo}{2007}]{lopez2007depth}
L{\'o}pez-Pintado, S. and J.~Romo (2007).
\newblock Depth-based inference for functional data.
\newblock {\em Computational Statistics \& Data Analysis\/}~{\em 51\/}(10),
  4957--4968.

\bibitem[\protect\citeauthoryear{Meiklejohn, Pomarole, Jordan, Levchenko,
  McCoy, Voelker, and Savage}{Meiklejohn et~al.}{2013}]{meiklejohn2013fistful}
Meiklejohn, S., M.~Pomarole, G.~Jordan, K.~Levchenko, D.~McCoy, G.~M. Voelker,
  and S.~Savage (2013).
\newblock A fistful of bitcoins: characterizing payments among men with no
  names.
\newblock In {\em Proceedings of the 2013 conference on Internet measurement
  conference}, pp.\  127--140.

\bibitem[\protect\citeauthoryear{M{\"u}ller, Stadtm{\"u}ller,
  et~al.}{M{\"u}ller et~al.}{2005}]{muller2005generalized}
M{\"u}ller, H.-G., U.~Stadtm{\"u}ller, et~al. (2005).
\newblock Generalized functional linear models.
\newblock {\em Annals of Statistics\/}~{\em 33\/}(2), 774--805.

\bibitem[\protect\citeauthoryear{Piotr, Hanny, Byeong, and Sangalli}{Piotr
  et~al.}{2017}]{piotr2017special}
Piotr, K., O.~Hanny, P.~Byeong, and L.~M. Sangalli (2017).
\newblock Special issue on functional data analysis.

\bibitem[\protect\citeauthoryear{Ramsay, Wickham, Ramsay, and deSolve}{Ramsay
  et~al.}{2020}]{ramsay2020package}
Ramsay, J., H.~Wickham, M.~J. Ramsay, and S.~deSolve (2020).
\newblock Package ‘fda’.

\bibitem[\protect\citeauthoryear{Ramsay and Silverman}{Ramsay and
  Silverman}{2005}]{RamsaySilverman2005book}
Ramsay, J.~O. and B.~W. Silverman (2005).
\newblock {\em Functional data analysis\/} (2nd ed.).
\newblock Springer.

\bibitem[\protect\citeauthoryear{Song, Deng, Lee, and Kwon}{Song
  et~al.}{2008}]{song2008optimal}
Song, J.~J., W.~Deng, H.-J. Lee, and D.~Kwon (2008).
\newblock Optimal classification for time-course gene expression data using
  functional data analysis.
\newblock {\em Computational biology and chemistry\/}~{\em 32\/}(6), 426--432.

\bibitem[\protect\citeauthoryear{Thompson}{Thompson}{1988}]{thompson1988point}
Thompson, W. (1988).
\newblock {\em Point process models with applications to safety and
  reliability}.
\newblock Springer Science \& Business Media.

\bibitem[\protect\citeauthoryear{Wang, Chiou, and M{\"u}ller}{Wang
  et~al.}{2016}]{wang2016functional}
Wang, J.-L., J.-M. Chiou, and H.-G. M{\"u}ller (2016).
\newblock Functional data analysis.
\newblock {\em Annual Review of Statistics and Its Application\/}~{\em 3},
  257--295.

\bibitem[\protect\citeauthoryear{Yao, M{\"u}ller, and Wang}{Yao
  et~al.}{2005}]{yao2005functional}
Yao, F., H.-G. M{\"u}ller, and J.-L. Wang (2005).
\newblock Functional data analysis for sparse longitudinal data.
\newblock {\em Journal of the American statistical association\/}~{\em
  100\/}(470), 577--590.

\end{thebibliography}
\end{document}